\documentclass[authoryear,preprint,review,12pt,fleqn]{elsarticle}
\usepackage[utf8]{inputenc}
\usepackage[a4paper, total={6in, 10in}]{geometry}
\usepackage{xcolor}
\usepackage{amsmath}
\usepackage{mathtools}
\usepackage{amssymb}
\usepackage{caption}
\usepackage{textcomp}
\usepackage{booktabs}
\usepackage{subcaption}
\usepackage{graphicx}
\usepackage{graphics}
\usepackage{soul}
\usepackage{multicol}
\usepackage{csquotes}
\usepackage{natbib}
\usepackage{float}
\usepackage{adjustbox}
\usepackage{tikz}
\usepackage{multirow}

\usepackage{lineno}

\journal{xx}

\begin{document}

\begin{frontmatter}



\title{Numerical Modeling of Effective Thermal Conductivity for Polymineralic Rocks using Lattice Element Method}


\author[1]{Nima Haghighat\corref{cor1}}
\ead{nima.haghighat@ifg.uni-kiel.de}
\author[1]{Amir S. Sattari}
\author[1]{Hem B. Motra}
\author[1]{Frank Wuttke}

\cortext[cor1]{Corresponding author}
\affiliation[1]{organization={Geomechanics \& Geotechnics Department, Kiel University},
            addressline={24118, Kiel}, 
            country={Germany}}

\begin{abstract}
Accurate prediction of rock thermal conductivity under in-situ conditions is essential for characterizing subsurface heat flow. This study presents a numerical framework based on the Lattice Element Method (LEM) for simulating the effective thermal conductivity of polymineralic rocks under coupled pressure–temperature conditions. The model resolves interactions among heat transfer, grain contacts, and mechanical deformation within a microstructure-representative lattice. The methodology enables consistent treatment of heat conduction, nonlinear contact evolution, and thermally induced intergranular fracturing. Heterogeneity is introduced through a stochastic, volume-fraction-constrained discretization that preserves the measured mineral composition and porosity, while mineral anisotropy and fracture behavior are captured through element-level constitutive laws. The framework is evaluated using experimental data for two dry sandstones under ambient and elevated pressures and temperatures. Effective thermal conductivity is computed over the same pressure–temperature ranges and compared directly with the measurements. The results indicate that the predictions are capable of reproducing the characteristic trends and absolute levels. The close agreement between experimental observations and model predictions confirms that the thermo-mechanical coupled LEM provides a physically grounded and transferable approach for modeling heat transport in heterogeneous, polymineralic media.

\end{abstract}


\end{frontmatter}


\section{Introduction}
Many geoengineering applications rely on accurate knowledge of the temperature and terrestrial heat flow within the Earth’s crust. This includes geothermal energy production, hydrocarbon exploration, underground waste storage, and subsurface energy storage systems, all of which require reliable estimates of the thermal regime at depth. Since direct measurements of temperature and heat flow are limited to shallow depths, modeling approaches are needed to extrapolate conditions into the deeper crust~\citep{seipold1995variation}. The accuracy of such models depends critically on the thermal transport parameters of rocks, with thermal conductivity being the most important property governing heat transfer in the subsurface~\citep{norden2020temperature}. Since rock is a multiphase system composed of a solid mineral framework and a pore-filling fluid, its thermal conductivity is generally expressed as an effective thermal conductivity (ETC). The ETC represents the combined contributions of the matrix thermal conductivity and the fluid thermal conductivity, governed by porosity and pore geometry~\citep{zimmerman1989thermal}. In addition to these intrinsic factors, external conditions such as pressure and temperature further modify the ETC~\citep{abdulagatov2006effect}. 

Thermal conductivity measurements of rocks can be performed either in the field or in the laboratory. However, laboratory-based methods are generally preferred due to their speed, cost-effectiveness, and better control over boundary conditions~\citep{barry2013thermal}. These methods are categorized into steady-state and transient techniques. Steady-state techniques measure heat flux once a constant temperature gradient is established, with commonly used methods including the guarded hot plate~\citep{astm1997177}, divided-bar~\citep{sass1971heat, sass1984thermal}, and radial heat flow techniques~\citep{cezairliyan2012compendium}. Transient methods, on the other hand, assess thermal conductivity by measuring a material's response to a pulse or periodic heat source. These methods include transient line source~\citep{von1959measurement, horai1971thermal}, transient plane source~\citep{gustafsson1991transient}, and laser flash techniques~\citep{parker1961flash}. A thorough comparison between mentioned methods can be found in \cite{zhao2016measurement} and \cite{kerschbaumer2019comparison}.

Beyond direct measurements, predictive approaches are particularly valuable, as thermal conductivity cannot be practically determined for every rock type under varying environmental conditions, such as changes in temperature, pressure, or fluid saturation. There is no universal agreement on how such predictive models should be classified. However, in the majority of the studies, these approaches have been categorized into three distinct groups of empirical models, mixing models, and theoretical models based on heat transfer mechanisms~\citep{midttemme1997thermal,dong2015critical,song2023influencing}.  While empirical models have proven useful in capturing certain datasets, they are inherently limited in scope. Their reliance on fitting parameters to specific rock types or experimental conditions restricts their predictive power when applied to lithologies or environments outside the calibration range, and hence are not discussed here. In contrast, mixing models basically describe the ETC as a function of the thermal conductivity of the constituent phases (solid matrix and pore fluid), weighted by porosity. Their popularity lies in their simplicity and the relatively small number of input parameters required. Theoretical models, on the other hand, are developed based on the fundamental mechanisms of heat transfer through simplified geometries. Both mixing and theoretical models have been widely applied to interpret and reproduce laboratory-measured thermal conductivities~\citep{sass1971thermal, zimmerman1989thermal,fuchs2013evaluation,chopra2018evaluate,tatar2021predictive,song2023influencing}. Nonetheless, as noted by \citet{abdulagatova2009effect}, the main challenge in constructing robust predictive models lies in capturing the complex and irregular geometries of rock microstructures. Such complexity makes it extremely difficult, if not impossible, to derive universally accurate theoretical predictions. Even with well-characterized microstructures, additional complications such as interfacial thermal resistances and inherent anisotropies further limit the reliability of these models. 

These limitations highlight the need for predictive approaches that incorporate more detailed representations of rock composition and structure. In this context, numerical modeling methods are becoming increasingly important, as they offer a promising way to simulate the thermal behavior of multiphase media. By building on established thermodynamic principles and heat transport mechanisms, these approaches can provide a more realistic framework for predicting rock thermal conductivity beyond the simplifying assumptions of conventional models~\citep{dongxing2021determination}. Many studies in the literature that estimate ETC of multi-phase material using numerical modeling follow a similar two-step workflow, where the micro-structural information of the porous medium is first acquired before the governing heat transport equations are solved locally within this digital representation to derive the effective thermal properties of the rock~\citep{wang2007mesoscopic1}. The microstructure of the studied sample can either be reconstructed using numerical reproduction algorithms or captured directly through digital imaging techniques. For example, \citet{wang2007mesoscopic1,wang2007mesoscopic2} employed the quartet structure generation set (QSGS) method, which is based on cluster growth theory, to reproduce randomly distributed multiphase granular porous media and investigate the ETC of isotropic porous systems. \citet{chen2015evaluation} investigated the ETC of asphalt concrete via Finite Element Modeling (FEM), where the heterogeneous microstructure was generated by randomly distributing aggregates of different sizes together with air voids within the asphalt binder. \cite{li2020meso} employed the Monte Carlo simulation to represent the random distribution of voids and then used FEM to solve for the mesoscale ETC of soil. To reconstruct the microstructure of various multiphase porous building materials, \citet{hussain2020thermal} applied the random generation of macro–meso pores (RGMMP) method, coupled with the lattice Boltzmann approach, to estimate their ETC. They concluded that the results obtained from the RGMMP algorithm were more accurate than those produced by the previously proposed QSGS method. \citet{janssen2022impact} investigated the influence of pore structure parameters on the thermal conductivity of porous building blocks using two synthetic pore structure generation algorithms, namely Random-Bubble-Insertion (RBI) and Watershed-Based (WSB). Their results highlighted the effects of structural characteristics such as total porosity, average pore size, and pore size distribution on the ETC. While reconstruction algorithms provide valuable insights by enabling rapid design and testing of new microstructures and by facilitating targeted studies of specific microscale parameters, digital imaging techniques allow the actual microstructure of a sample to be incorporated directly. These methods can resolve features down to the micrometer scale, making it possible to capture actual pore connectivity, grain–pore contact areas, and whether phases form continuous or dispersed networks. Examples of digital image–based approaches applied to thermal conductivity estimation can be found in \citet{dongxing2021determination}, \citet{najafi2023assessment}, and \citet{caniato2024modelling}, among others.

Most of the studies discussed above treat heat transfer in isolation, without coupling to mechanics. Yet, as noted earlier, pressure and temperature invariably coexist at depth and both influence ETC. Predictive models should therefore account for thermo–mechanical interactions: purely thermal formulations cannot directly capture pressure effects and they overlook processes such as thermally induced cracking. Such cracking can arise from mismatches in thermal expansion coefficients among minerals in polymineralic rocks and can markedly reduce effective thermal conductivity. In this regard, \citet{shen2015mesoscopic,shen2017experimental} numerically investigated cracking effects on the ETC of concrete, modeling a three-phase composite comprising a mortar matrix, randomly distributed aggregates, and an interfacial transition zone (ITZ). They reported a sharp reduction in ETC associated with microcrack development in both the mortar and the ITZ. However, in their framework, cracks were introduced by applying a purely mechanical load, and the resulting damage was then mapped into the thermal analysis by redefining the mortar’s thermal conductivity. Although thermo–mechanically coupled models have been explored in various numerical frameworks, only a few have quantified how microcrack formation, pore-structure alteration, or mineral-scale thermal mismatch influence heat transport in geomaterials. This limited understanding highlights the need for modeling approaches capable of simultaneously resolving the mechanical and thermal responses of heterogeneous rocks while preserving microstructural detail. The Lattice Element Method (LEM) provides such a framework. At the mesoscale, LEM represents the medium as a discrete network of springs, trusses, or beam elements, enabling straightforward assignment of phase-specific properties and seamless incorporation of mineralogical and pore-scale heterogeneities through elementwise parameterization. In recent studies by our group, the LEM was generalized to include thermo-mechanical coupling, enabling reproduction of heat transfer in heterogeneous materials and accurate estimation of effective thermal conductivity relative to experimental data~\citep{sattari2017meso, rizvi2018numerical, rizvi2020soft}. However, explicit micromechanical fracturing has not been represented, limiting the ability to capture pressure- and temperature-induced damage and its influence on thermal transport processes. The present work addresses this gap by targeting polymineralic rocks and developing a formulation that further resolves intergranular fracture, grain-contact quality, and anisotropic mineral conductivities. The structure of the study is as follows. First, the modeling foundations are introduced, covering the thermal model, the thermo-mechanical coupling, and the fracture initiation and growth criteria. The results then open with a brief overview of the studied samples, followed by the reference experimental measurements, the room-temperature verification with comparisons to mixing models, and finally the verification under in-situ conditions across the pressure–temperature grid.

\section{Methodology}

The details of the proposed thermo-mechanical coupled LEM are presented in this section. The overall methodology can be divided into three parts: (1) representation of the domain with LEM, (2) the mechanical formulation (spring-network), and (3) the thermal formulation. In the present work, only the domain representation and the thermal formulation are discussed. The mechanical aspects are not addressed here, and further details can be found in \citep{ThesisSattari, sattari2017meso}.

\subsection{Domain representation using LEM}
\begin{figure}[h!]
  \centering
  \begin{subfigure}[b]{0.32\textwidth}
    \includegraphics[width=\textwidth]{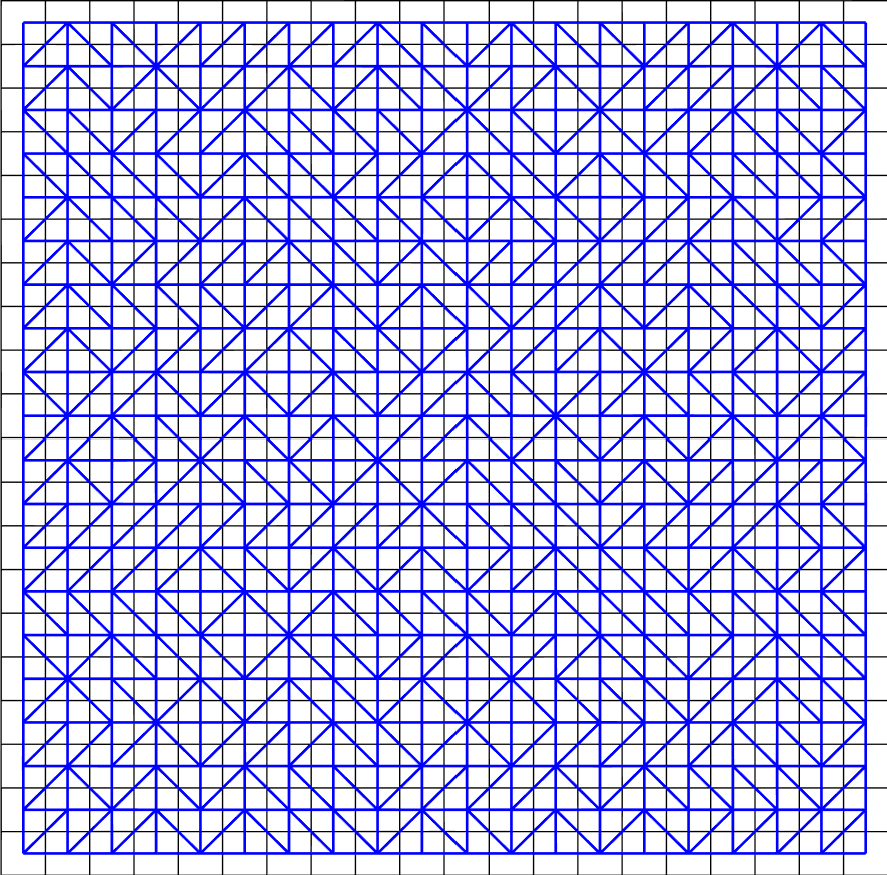}
    \subcaption{}
  \end{subfigure}
  \begin{subfigure}[b]{0.32\textwidth}
    \includegraphics[width=\textwidth]{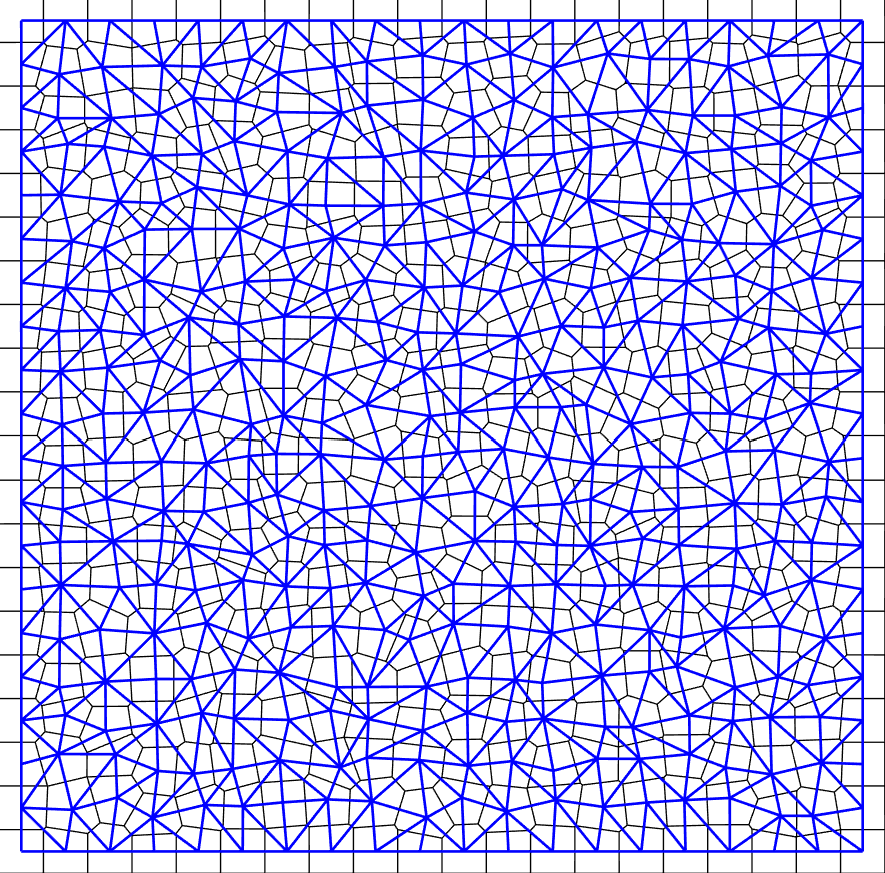}
    \subcaption{}
  \end{subfigure}
  \begin{subfigure}[b]{0.32\textwidth}
    \includegraphics[width=\textwidth]{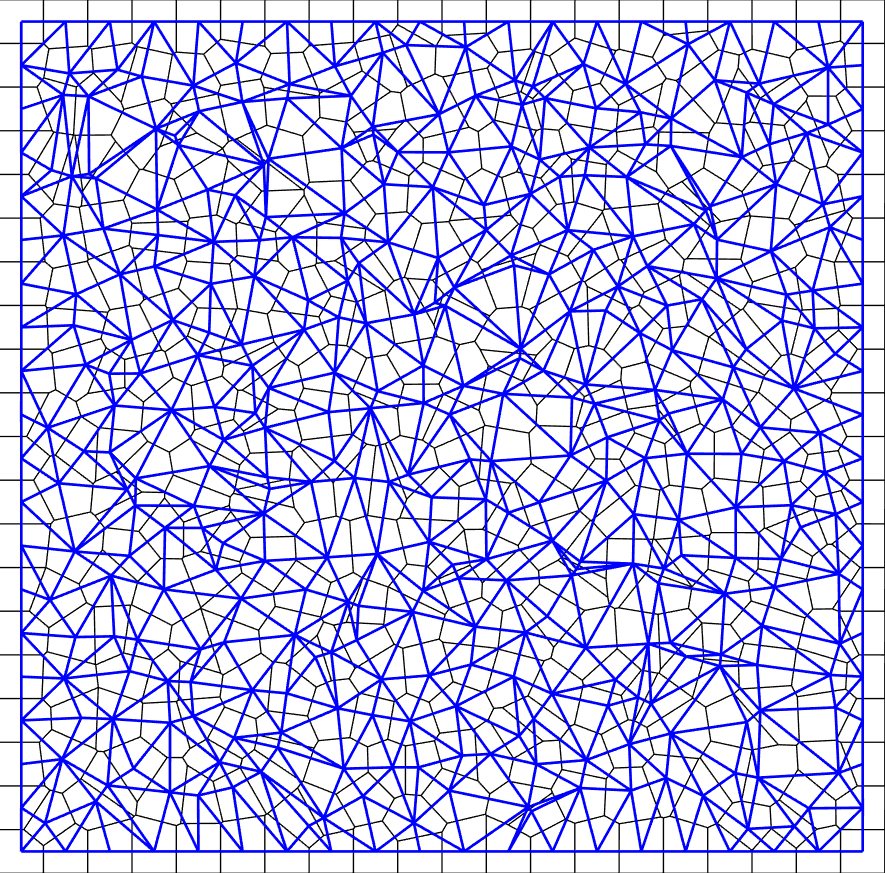}
    \subcaption{}
  \end{subfigure}
  \caption{LEM discretizations generated with increasing geometric randomness $\text{R}_{\text{F}}$: (a) $\text{R}_{\text{F}} \approx 0$, (b) $\text{R}_{\text{F}} = 0.5$, and (c) $\text{R}_{\text{F}}=1$. The blue lines represent the Delaunay edges forming the lattice of intergranular contacts, which transmit mechanical loads and heat.\label{fig:mesh_alpha}}
\end{figure}

In the proposed LEM framework, the particle arrangement and their contacts are derived from a DEM-based geometry, which is discretized using a vectorizable random lattice. The procedure relies on Voronoi tessellation, where nodal points are generated within subdivided squares controlled by the parameter $\text{R}_{\text{F}}$, which defines the degree of randomness. Higher values of $\text{R}_{\text{F}}$ lead to greater irregularity in the lattice, thereby introducing micro-scale heterogeneity that influences the meso- and macro-scale response of the medium. Auxiliary points are employed to ensure well-defined cells at the boundaries. Once the Voronoi diagram is constructed, the contact network and contact lengths are obtained by Delaunay triangulation. Examples of the resulting discretizations for different values of the randomness parameter $\text{R}_{\text{F}}$ are shown in Fig.~\ref{fig:mesh_alpha}. Increasing $\text{R}_{\text{F}}$ produces a more irregular lattice with a wider distribution of contact lengths, whereas a smaller $\text{R}_{\text{F}}$ yields a more uniform tessellation.

In this regard, material heterogeneity can be incorporated into the calculations through different strategies. In the present work, heterogeneity is introduced by means of a particle overlay. Depending on the chosen representation, a single Voronoi cell may correspond to an individual particle, or multiple cells may be grouped to form a single particle. The particle overlay then defines the thermo-mechanical properties assigned to the lattice elements. Three approaches can be employed to generate such overlays: (1) random assignment of each cell to a material, (2) particle packings constructed according to a particle size distribution (PSD)~\cite{}, or (3) image-based processing.

\subsection{Thermal model}
In the discrete element methods, two principal approaches have been developed to simulate heat transfer in granular media. The first defines the contact length from the geometric relation between particle radii and contact angle, independent of contact forces~\citep{feng2008discrete}. The second, based on Hertzian contact theory, relates the thermal conductance between particles to the normal contact force and the elastic properties of the grains~\citep{zhang2011study}. In this study, the Hertzian formulation is adopted in the LEM framework in order to capture the influence of contact forces on heat transfer, while assuming smooth particle–particle contact surfaces.

\begin{figure}
    \centering
    \includegraphics[width=0.3\linewidth]{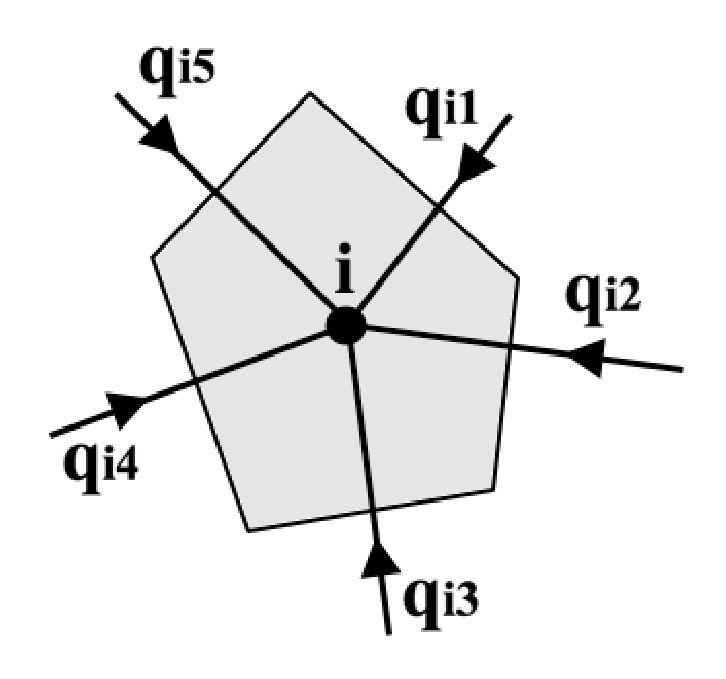}
    \caption{A schematic representation of particle i and the thermal fluxes exchanged with its neighboring particles.}
    \label{fig:LEM_Thermal_Steady}
\end{figure}

The governing heat conduction equation for particle $i$ can be written as
\begin{equation}
    \rho_i c_i \nu_i \frac{dT_i}{dt}
    - \nabla \cdot \left( k \nabla T_i \right)
    - \rho_i \dot{q}_i = 0
    \label{eq:heat_equation}
\end{equation}

\noindent where $\rho_i$ is the density, $c_i$ is the specific heat capacity, $\nu_i$ is the volume, $t$ is the time, $T_i$ is the temperature, $k$ is the thermal conductivity, and $\dot{q}_i$ is the heat generation density of particle $i$. Under steady-state conditions and in the absence of internal heat sources, \emph{i.e.}, $\dot{q}_i(t) = 0$,  
Eq.~(\ref{eq:heat_equation}) reduces to

\begin{equation}
    \nabla \cdot \left( k \nabla T_i \right) 
    = \sum_{j=1}^{N_c} q_{ij} = 0 ,
    \label{eq:heat_equation_summ}
\end{equation}

\noindent where $q_{ij}$ denotes the heat exchange between particle $i$ and its $j$-th neighboring particle, and $N_c$ is the number of contacts of particle $i$.  
Eq.~(\ref{eq:heat_equation_summ}) therefore states that, at equilibrium, the net heat inflow and outflow at each particle balance to zero. This principle is illustrated schematically in Fig.~\ref{fig:LEM_Thermal_Steady}, which shows the heat fluxes entering a representative particle from its neighbors. To evaluate the individual terms in Eq.~(\ref{eq:heat_equation_summ}), the heat flux exchanged between a pair of contacting particles must be considered.  
For a contact between particles $i$ and $j$, the exchange is written as

\begin{equation}
    q_{ij} = \alpha_{Rc}\,h_{ij}\,(T_i - T_j) ,
    \label{eq:qij}
\end{equation}

\noindent where $\alpha_{Rc}$ is the contact quality factor between two minerals, $h_{ij}$ denotes the effective thermal conductance of the contact, and $T_i$ and $T_j$ are the particle temperatures. The contact quality factor represents the amount of heat loss during heat transmission between two nonidentical minerals and is in inverse correlation with the thermal contact resistance ($\alpha_{Rc} \sim \frac{1}{Rc}$), varying between 0 to 1. Thermal resistance arises from microscopic surface roughness, contact imperfections, and differences in thermal properties, leading to a temperature discontinuity at the interface. In this study, $h_{ij}$ is obtained from a modified analytical expression based on Hertzian contact theory under the assumption of perfectly smooth contact surfaces. In this regard, the inter-particle thermal conductivity can be written as  

\begin{equation}
    h_{ij} = k \left( A \pm 2 \left( \frac{3 F_{n} L}{4 E_b} \right)^{\tfrac{1}{3}} \right) ,
    \label{eq:hij}
\end{equation}

\noindent where $\text{A}$ is the cross-sectional length, $\text{L}$ is the lattice element length, $F_{\text{n}}^{\text{c}}$ is the normal contact force between two neighboring particles, $\text{E}_{\text{b}}$ is the effective elastic modulus, and $\text{k}$ is the thermal conductivity.

\begin{figure}[htbp]
    \centering
    \begin{subfigure}[t]{0.32\textwidth}
        \centering
        \includegraphics[scale=0.35]{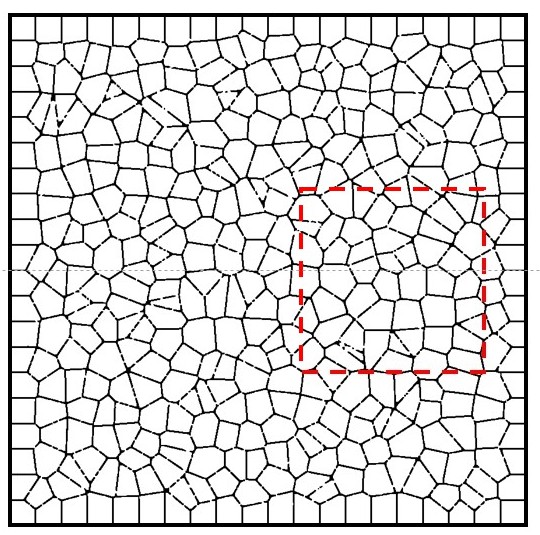}
        \subcaption{}
    \end{subfigure}
    \begin{subfigure}[t]{0.32\textwidth}
        \centering
        \includegraphics[width=\textwidth]{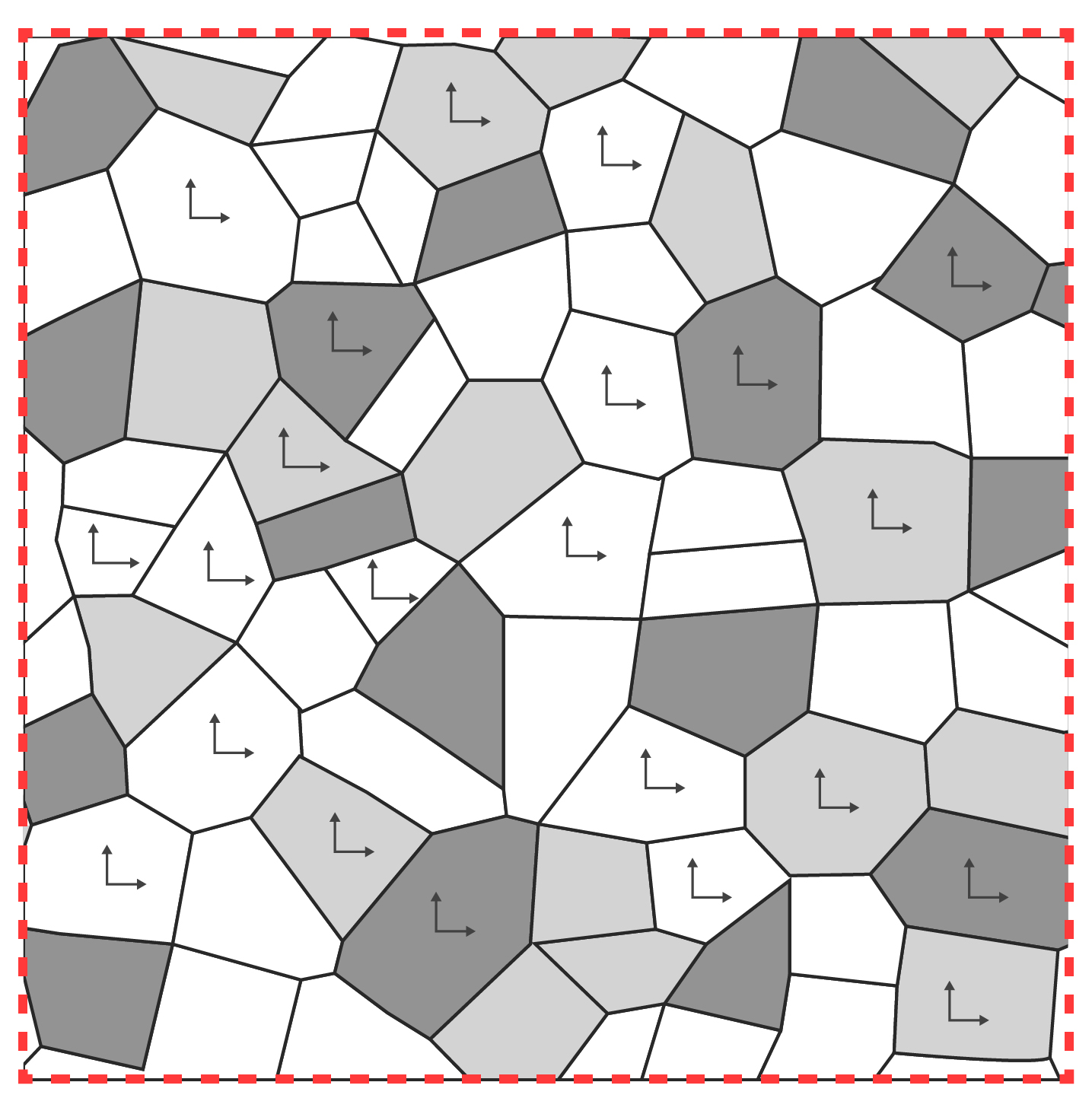}
        \subcaption{}
    \end{subfigure}
    \begin{subfigure}[t]{0.32\textwidth}
        \centering
        \includegraphics[width=\textwidth]{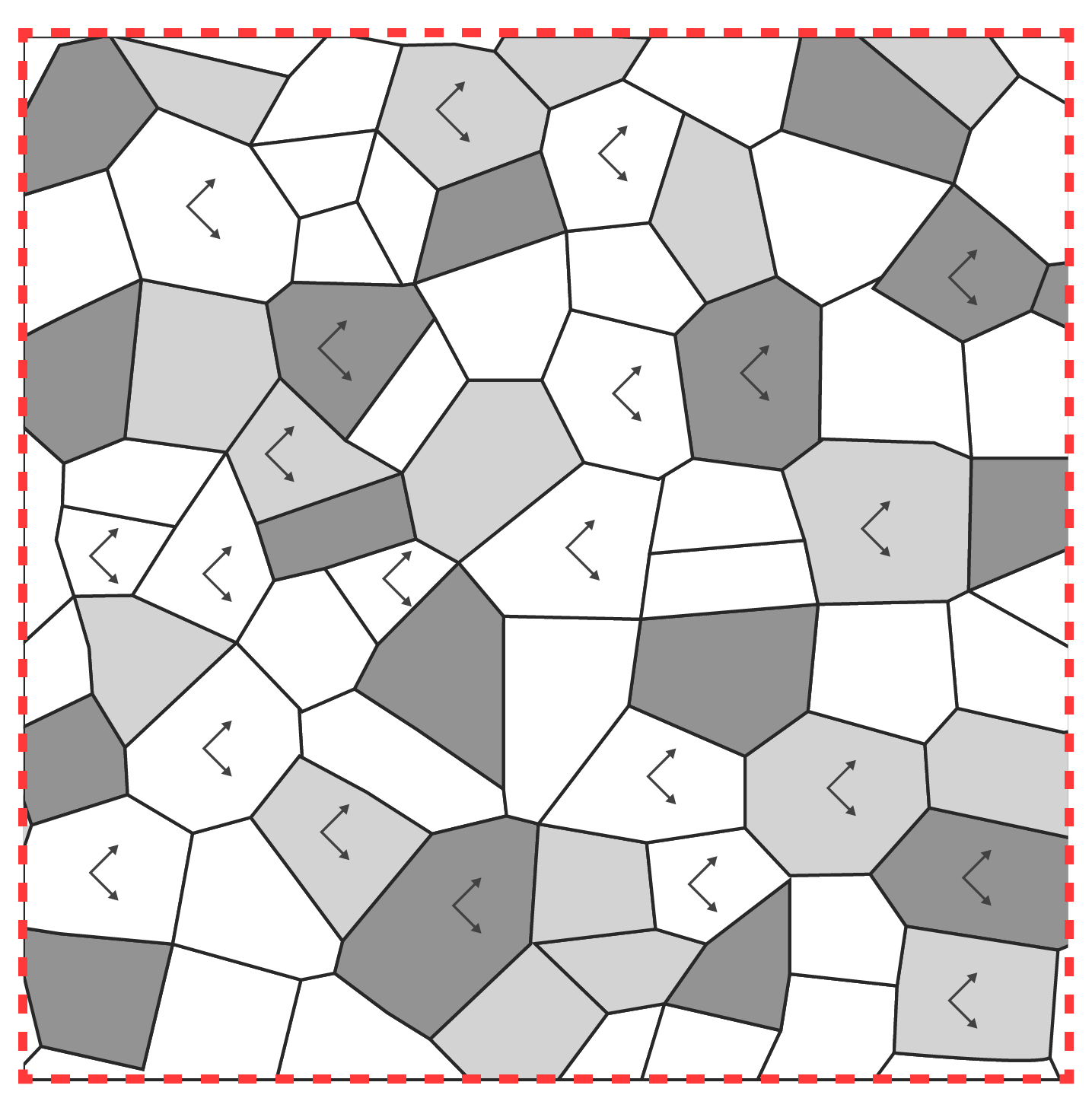}
        \subcaption{}
    \end{subfigure}\caption{(a) Example of the discretized lattice and selected domain; assigned mineral grains with crystallographic axes defining thermal anisotropy with $\text{k}_{\parallel}$ (b) aligned with the vertical heat-flow direction, and (c) a configuration rotated by $45~^{\circ}$ .\label{fig:rotation_mineral_conductivity}}   
\end{figure}

In the LEM framework, mineral anisotropy is represented by assigning each grain a local coordinate system aligned with its crystallographic axes. The direction of the c-axis defines the principal direction of anisotropy, with $\text{k}_{\parallel}$ denoting the conductivity along the c-axis and $\text{k}_{\perp}$ the conductivity in the basal plane~\citep{birch1940thermal}. For the simulations presented here, all grains were assigned the same orientation, with the c-axis assumed parallel to the vertical direction of heat flow. The influence of grain orientation was later assessed through a sensitivity analysis by systematically rotating the c-axis from $0~^{\circ}$ (vertical) to $90~^{\circ}$ (horizontal). An example of the discretized domain and the corresponding orientations is shown in Fig.~\ref{fig:rotation_mineral_conductivity}.

In this regard, the thermal conductivity value in Eq.~(\ref{eq:hij}) is calculated based on the difference between orientation of a lattice element ($\mathrm{\theta}$) and the defined crystallographic axes ($\mathrm{\theta}_{\text{c}}$) as,

\begin{equation}
    k =  k_{\parallel}\cos\left(\theta-\theta_c\right)^2+k_{\perp}\sin\left(\theta-\theta_c\right)^2.
\end{equation}

The complete thermal problem is then obtained by assembling Eq.~\ref{eq:qij} for all particles in the domain, leading to a system of linear algebraic equations.  
This system is solved numerically to determine the temperature distribution in the granular assembly. Eventually, the resultant normal force of thermal expansion ($F_n^t$) is equal to

\begin{equation}
\label{eq:therm_strain3}
F_n^t=k_n u_t = E A\varepsilon_t = E A \alpha \Delta T,
\end{equation}

where $\alpha$ is the thermal expansion coefficient, and $\Delta T$ is the temperature change across two consecutive iterations. The calculated thermal expansion forces are transformed into the mechanical solver to analyze the initiation and propagation of the fracture in the domain. In the current methodology, a toughness-based rule compares the local driving force to a critical value so that bond failure and crack extension occur only when this value is exceeded.  In the case of an elastic-brittle Mode I spring bond, the total strain energy is released when the element breaks. Hence,

\begin{equation}
\label{eq:mech_failure_3}
\frac{1}{2\cdot}k_n\cdot u^2=G_\text{Ic}\cdot A    \qquad  \rightarrow \qquad
\sigma_t=\sqrt{\frac{2G_\text{Ic}E}{L_c}}=K_\text{Ic}\sqrt{\frac{2E}{L_cE^*}},
\end{equation}

where $L_c$ is the rock's characteristic length scale, and $E^*$ is the effective Young's modulus, which depends on the adopted 2D approximation rule. 

\section{Results}

The benchmarking of the LEM framework, together with a comprehensive sensitivity analysis of key model parameters, has been reported previously. In particular, the influence of the randomness parameter $\text{R}_{\text{F}}$ on the effective thermal conductivity was examined in detail, establishing the robustness of the approach and clarifying the role of geometric irregularity~\citep{sattari2017meso}. These elements are not repeated here. Building on that foundation, the present work develops a thermomechanical extension for polymineralic rock, with special consideration given to intergranular fracturing via a toughness-based criterion.

The results are organized in three parts, with two dry sandstone samples, S\#8 and S\#12, serving as the common reference across all comparisons. First, the experimental thermal conductivity data for S\#8 and S\#12 are presented to establish the empirical baseline. Second, the ability of the developed methodology to reproduce bulk thermal conductivity under ambient conditions is examined using measurements at room temperature and atmospheric pressure for these samples, accompanied by a comparison to well-known mixing models. Third, the analysis is extended to in-situ relevant conditions, where LEM predictions for the thermal conductivity of S\#8 and S\#12 are evaluated over varying confining pressure and temperature to assess the predictive capability of the approach. Key descriptors of S\#8 and S\#12 are summarized below.

The rock specimens were extracted from sandstone outcrops sampled from the Frosinone Formation in the central Apennines, Italy~\citep{smeraglia2014tectonic}. The primary distinction between these two sandstone types lies in their porosity values and the distance from the nearby fault where they were collected~\citep{smeraglia2014tectonic}. Sandstone S\#8 has a porosity of 8.4~\%, while S\#12 has a porosity of 5.68~\%. The XRD results, further backed by the XRF analysis, revealed that quartz is the dominant constituent mineral in the studied sample, with other minerals including dolomite, albite, and calcite. Rietveld analysis for S\#8 shows that quartz, calcite, dolomite, and albite constitute approximately 40\%, 10\%, 16\%, and 32\% of the rock volume, respectively. Similarly, S\#12 consists of about 37\% quartz, 19\% calcite, 16\% dolomite, and 26\% albite.

\begin{figure}[htbp]
    \centering
    \begin{subfigure}[t]{0.35\textwidth}
        \centering
        \includegraphics[width=\textwidth]{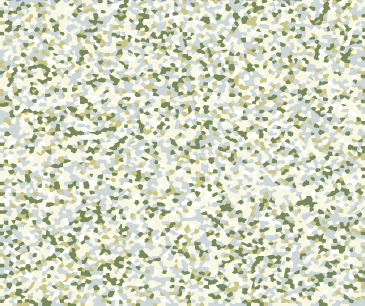}
        \subcaption{\label{fig:S8_LEM}}
    \end{subfigure}
    \begin{subfigure}[t]{0.35\textwidth}
        \centering
        \includegraphics[width=\textwidth]{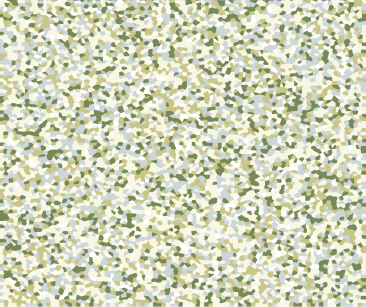}
        \subcaption{\label{fig:S12_LEM}}
    \end{subfigure}
    \begin{subfigure}[t]{0.2\textwidth}
        \centering
        \includegraphics[width=\textwidth]{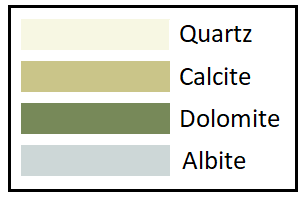}
    \end{subfigure}
    \caption{Stochastic LEM discretizations of (a) S\#8, and (b) S\#12 based on measured phase fractions. Phases are randomly assigned according to the experimentally determined proportions.}
\end{figure}

As mentioned previously, the material heterogeneity can be incorporated into the calculations either through random assignment or through image-based processing. Initially, both routes were explored. Computed tomography was attempted first to obtain specimen-scale maps suitable for discretization. However, the available setup did not resolve the pore scale characteristic of these sandstones, and attenuation contrasts between common mineral phases were insufficient for reliable segmentation. As a result, neither porosity nor phase topology could be mapped with confidence from the CT volumes. Thin-section imaging was then pursued to provide mineral scale detail. At the field of view required to capture representative textures, phase boundaries could not be delineated consistently, and intensity overlap hindered robust automated classification. Given these constraints, a stochastic representation was adopted. Phases were assigned by random tessellation conditioned on the measured volume fractions for each sample, and bulk porosity was imposed to match the independently determined value. The resulting discretizations are illustrated in Fig.~\ref{fig:S8_LEM} for S\#8 and Fig.\ref{fig:S12_LEM} for S\#12.

\subsection{Experimental Thermal Conductivity Results}
The thermal conductivity measurements at room temperature were performed using the Thermtest TLS-100 device, which operates based on the transient line heat source method. For each sample, an axial bore was drilled to accommodate the TLS-100 needle probe, with the bore diameter matched to the probe and sufficient clearance to all external surfaces. To minimize contact resistance and eliminate air gaps, a thin layer of thermal paste was applied within the bore before probe insertion. The average thermal conductivity, based on five repeated measurements, was found to be 2.11~$\text{W}~\text{m}^{-1}~\text{K}^{-1}$ for sample S8 and 2.5 ~$\text{W}~\text{m}^{-1}~\text{K}^{-1}$ for sample S12 (Table~\ref{tab:Transient_RoomTemp}).

\begin{table}[h!]
\centering
\caption{Room-temperature thermal conductivity values (~$\text{W}~\text{m}^{-1}~\text{K}^{-1}$) by TLS-100}
\begin{tabular}{|c| c c c c c|c|}
\hline
\textbf{Sample} & \textbf{1} & \textbf{2} & \textbf{3} & \textbf{4} & \textbf{5} & \textbf{average} \\
\hline
S8 & 2.02& 2.14& 2.13&2.16 & 2.12 & 2.11\\
S12 & 2.40 & 2.50 & 2.48& 2.61 & 2.52 & 2.50 \\
\hline
\end{tabular}
\label{tab:Transient_RoomTemp}
\end{table}

\begin{figure}[htbp]
    \centering
    \begin{subfigure}[t]{0.45\textwidth}
        \centering
        \includegraphics[width=\textwidth]{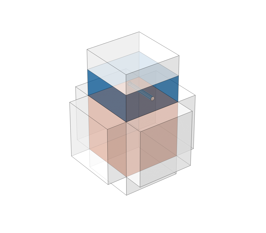}
        \subcaption{}
    \end{subfigure}
    \begin{subfigure}[t]{0.45\textwidth}
        \centering
        \includegraphics[width=\textwidth]{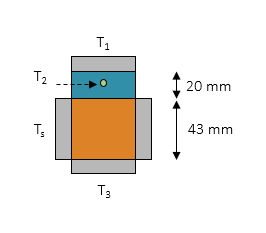}
        \subcaption{}
    \end{subfigure}
    \caption{(a) Three-dimensional schematic representation, and (b) cross-sectional view of the cubic press assembly of comparative steady state configuration. The blue block denotes the zirconia reference (axial thickness 20 mm) and the orange block denotes the rock sample (axial thickness 43 mm). The grey regions represent the top and bottom platens and the lateral guards held at temperatures $\text{T}_1$, $\text{T}_3$, and $\text{T}_{\text{S}}$, respectively. A constant axial gradient is imposed between $\text{T}_1$, and $\text{T}_3$ ($\text{T}_1 > \text{T}_{\text{S}}>\text{T}_3$). $\text{T}_2$ marks the measurement point, centered at the reference material height.}
    \label{fig:WP_TC_Schematic}
\end{figure}

As outlined in the introduction, thermal conductivity under in situ–relevant pressure and temperature was measured using a cubic press adapted for thermal testing. The apparatus applies confining pressures up to 600 MPa and temperatures up to 600 °C. Thermal conductivity was obtained by a comparative steady state method, in which the temperature drop across the specimen is compared with that across a reference of known conductivity. A zirconia plate served as the reference and was placed directly above the sample within the stack. An axial thermal gradient was imposed by independently regulating the top and bottom pistons, while the lateral pistons acted as guard heaters to reduce radial heat losses. The temperature drop within the reference was monitored at its midplane using a resistance temperature sensor. A schematic of the assembly is shown in Fig.~\ref{fig:WP_TC_Schematic}. Detailed aspects of the implementation, verification, and calibration are reported in the companion experimental study and are not repeated here~\citep{haghighat2025thermalacousticalmechanicalcharacterization}.

\begin{figure}[htbp]
    \centering
    \begin{subfigure}[t]{0.4\textwidth}
        \centering
        \includegraphics[width=\textwidth]{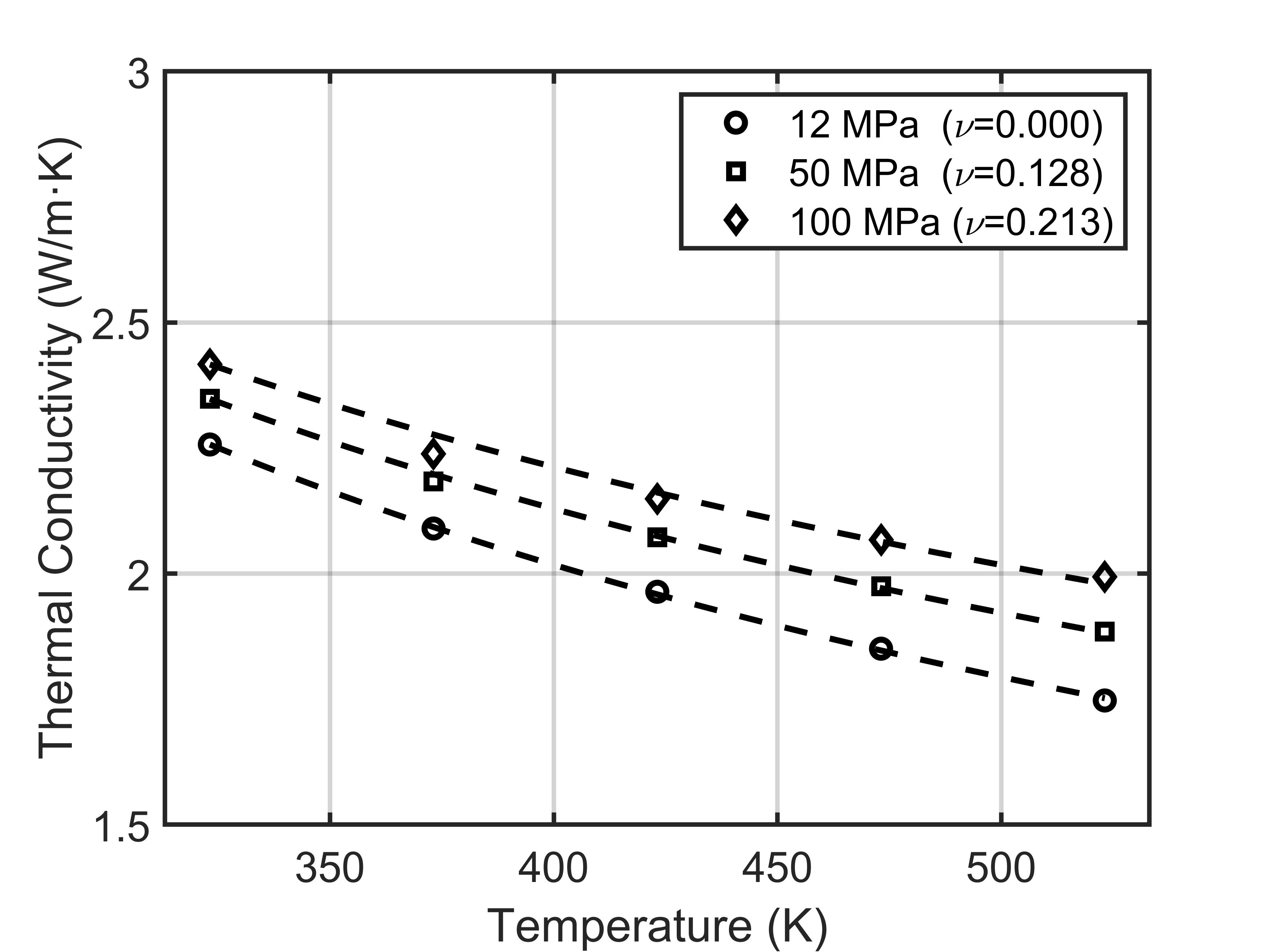}
        \subcaption{\label{fig:TC_S8}}
    \end{subfigure}
    \begin{subfigure}[t]{0.4\textwidth}
        \centering
        \includegraphics[width=\textwidth]{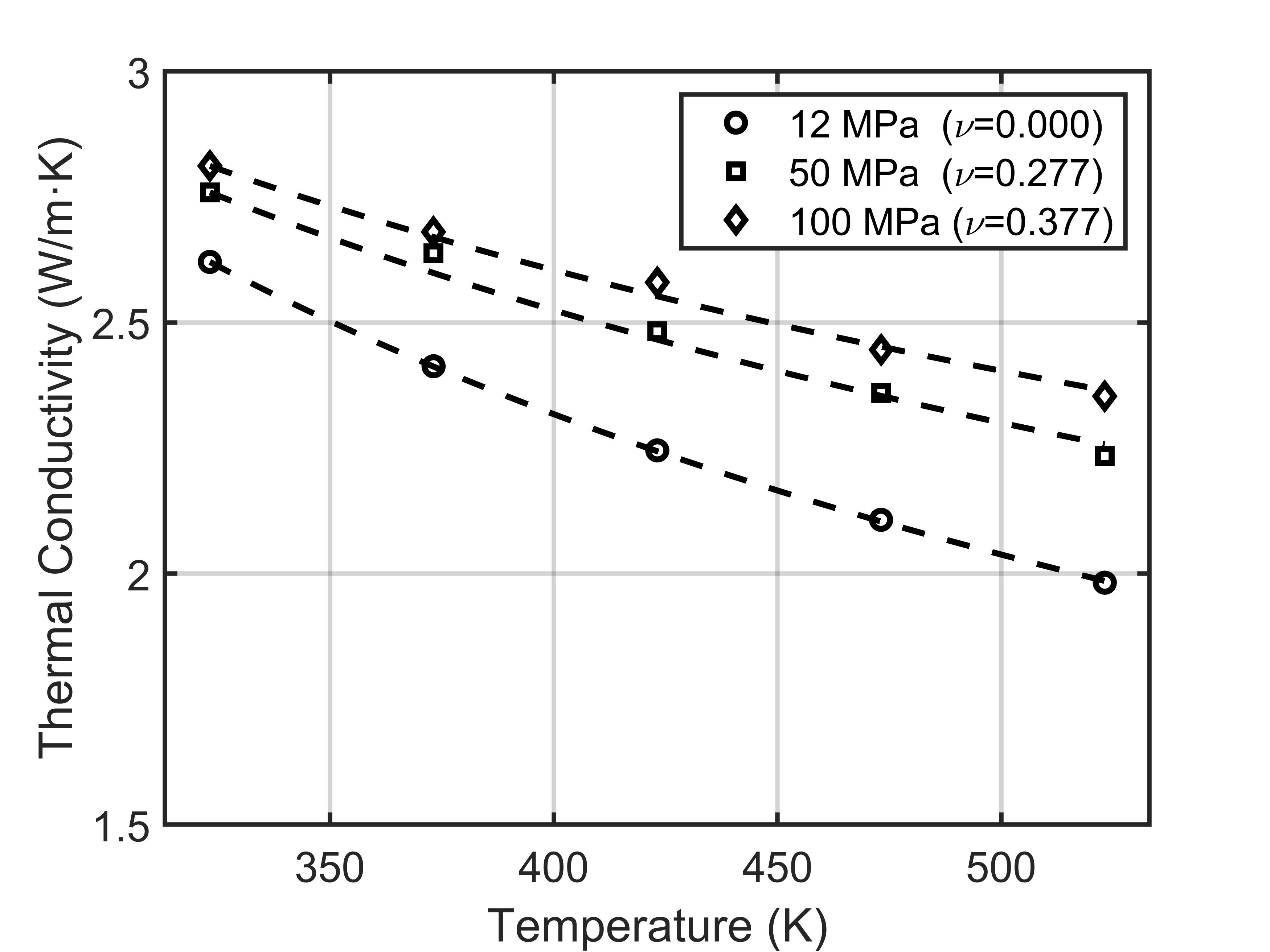}
        \subcaption{\label{fig:TC_S12}}
    \end{subfigure}
    \caption{Thermal conductivity of samples S\#8 (left) and S\#12 (right) as a function of temperature, evaluated at different confining pressures.}
    \label{fig:TC_s12_s8}
\end{figure}

Measurements in the cubic press were conducted at confining pressures of 12, 50, and 100 MPa. For each pressure level, temperature was increased from $50~^{\circ}\text{C}$ to $250~^{\circ}\text{C}$ in $50~^{\circ}\text{C}$ increments. According to \cite{emirov2021studies}, the combined effect of pressure and temperature on the thermal conductivity of a polymineralic rock can be expressed using the following semi-empirical relationship as

\begin{equation}
    k(T,P) = k(T_0,P).(T/T_0)^{n_0.(1-\nu(p))}
    \label{eq:lambda_TP}
\end{equation}

where, theoretically, $\text{n}_0$ defines the temperature-dependence of thermal conductivity at zero confining pressure, and $\nu(p)$ is the relative pressure coefficient of the exponent $n$. The obtained thermal conductivity values are shown in Fig.~\ref{fig:TC_S8} for S\#8 and Fig.~\ref{fig:TC_S12} for S\#12 along with the fitted curves using Eq.~(\ref{eq:lambda_TP}).  Here, $\text{n}_0$, defines the temperature dependence $\text{k}/\text{k}_0 = (\text{T}/\text{T}_0)^{\text{n}_0}$, at 12 Mpa of applied pressure. This relation, beyond the direct effect of pressure and temperature on thermal conductivity, captures the gradual reduction in temperature sensitivity with increasing confining pressure.

\subsection{Verification under atmospheric conditions}

In this section, the experimental results from thermal conductivity measurements at room temperature and atmospheric conditions are compared against the values predicted by established predictive models available in the literature, as well as the LEM results. Here, mixing models are adopted as a practical, physics-based baseline that is widely used to reproduce laboratory thermal conductivities with reasonable accuracy while requiring only a small number of input parameters. These models provide reference estimates or bounds for idealized material conditions, serving as useful benchmarks for interpreting laboratory data. In this context, five widely used fundamental structural models are selected to evaluate the capability of mixing models in predicting thermal conductivity at room temperature: the series and parallel models, the Hashin–Shtrikman bounds, and the effective medium theory (EMT). A comprehensive list of mixing models can be found in~\cite{abdulagatova2009effect}.

Assuming a two-component system, the ETC is expected to fall between the thermal conductivities of the solid matrix and the pore fluid. The so-called parallel and series models are the simplest forms of weighted averaging used to estimate the upper and lower bounds of ETC based on the thermal conductivities of the individual phases. In these models, the ETC is given by:

\begin{equation}
    k=\phi k_f + (1-\phi)k_s
\end{equation}
for the parallel model and by
\begin{equation}
    \frac{1}{k} = \frac{\phi}{k_f} + \frac{1-\phi}{k_s}
\end{equation} for the series model, where $\phi$ represents the porosity, $\text{k}_\text{f}$ the thermal conductivity of the pore fluid, and $\text{k}_\text{s}$ the conductivity of the solid phase. These two expressions were first introduced by~\cite{wiener1912theorie}, and are now widely recognized as representing the theoretical upper and lower bounds of effective conductivity in two-phase systems. These two bounds are often too far apart to provide reliable predictions for most geological materials. Moreover, they are considered physically unrealistic because they assume a body composed of alternating, perfectly aligned slabs of solid and fluid phases~\citep{zimmerman1989thermal}. To address these limitations, \cite{hashin1962variational} introduced a model based on variational principles, which significantly narrows the gap between the Wiener bounds and offers a more realistic estimation of effective thermal conductivity. Based on their model, the ETC can fall between 
\begin{equation}
    k_f + \frac{3k_f(k_s-k_f)(1-\phi)}{3k_f+(k_s-k_f)\phi}
\end{equation}
and
\begin{equation}
    k_s +\frac{3k_s(k_f-k_s)\phi}{3k_s+(k_f-k_s)(1-\phi)}
\end{equation} Despite being derived from different theoretical foundations, the Hashin–Shtrikman bounds are mathematically equivalent to the classical Maxwell–Eucken formulations~\citep{carson2005thermal}. These bounds can be physically realized by a space-filling assemblage of composite spherical shells, each of which having the same volume fraction of solid and fluid. Another widely applied theoretical approach is the EMT theory, which provides a self-consistent estimate of the effective thermal conductivity for composite materials~\citep{landauer1952electrical}. Unlike models that assume either a continuous matrix or isolated inclusions, EMT assumes a completely random distribution of components, without bias toward either phase. The effective thermal conductivity for such a structure is given by the EMT equation:
\begin{equation}
    (1-\phi)\frac{k_s - k}{k_s + 2k}+\phi\frac{k_f-k}{k_f+2k} = 0
\end{equation}

In addition to these individual models, researchers have, over the years, systematically combined the aforementioned basic formulations to develop hybrid or modified models tailored to specific applications. For example, \cite{pabst2019describing}, proposed a unifying fixed parameter equation as
\begin{equation}
     k=(1-\delta_{0N})\left[(1-f)(k^U)^N+f(k^L)^N\right]^{1/N}+\delta_{0N}\left[ exp[(1-f)ln(k^U)+f~ln(k^L)]\right]
     \label{eq:combi}
\end{equation}

where $\text{f}$ is the distribution factor, and $\delta_{0\text{N}}$ is the Kronecker delta, defined as $\delta_{0\text{N}} = 1$ for $\text{N} = 0$ and $\delta_{0\text{N}} = 0$ for $\text{N} \ne 0$. The parameter $\text{N}$ specifies the type of weighted mean used, such as harmonic, arithmetic, or geometric, for the combination of the models. $\text{k}^{\text{U}}$ and $\text{k}^{\text{L}}$ denote the upper and lower bounds of the basic structural model sets, including the series and parallel models, as well as the Hashin–Shtrikman (HS) bounds. 

Having reviewed the fundamental predictive models, the following section presents a comparison between the thermal conductivity of the studied sandstone samples at room temperature and the values estimated using these models. In order to compare these values with the theoretical bounds, the thermal conductivities of the fluid and solid phases, $k_f$ and $k_s$, must be defined. Since the samples were dry during the measurements, air is considered the pore fluid, and its thermal conductivity is taken as $k_f=0.026~\text{W}~\text{m}^{-1}~\text{K}^{-1}$. The matrix (solid phase) thermal conductivity is typically estimated based on the rock’s mineralogical composition. The geometric mean model is one of the most widely used approaches for this purpose, as it accounts for the volumetric fraction of each mineral phase and their individual thermal conductivities~\citep{midttemme1997thermal}:
\begin{equation}
    K_m =  \prod_{i=1}^{n} K_i^{v_i}
\end{equation}
where $\text{K}_i$ is the thermal conductivity of each mineral, and $\text{v}_i$ is the volume fraction of that mineral. The reported thermal conductivity of the rock-forming minerals is summarized in Table~\ref{tab:mineral_tc}. 
 
\renewcommand{\arraystretch}{1.5} 
\begin{table}[htbp!]
\centering
\caption{Temperature dependence of mineral thermal conductivity; [\citenum{birch1940thermal}]~:~\cite{birch1940thermal},~[\citenum{xiong2021thermal}]~:~\cite{xiong2021thermal}; The values superscripted with * are inerpolated}

\begin{tabular}{|l|c *{6}{c} c c|}
\hline
\multirow{2}{*}{\textbf{Mineral}} & 
\multicolumn{7}{r}{\textbf{Thermal Conductivity (W/m·K)}} &
&
\multirow{2}{*}{\textbf{Ref.}} \\
\cline{3-8}
 & \textbf{Temp. (\textdegree C):} & 25–35 & 50 & 100 & 150 & 200 & 250 & \multicolumn{2}{c|}{}\\
\hline
$\alpha$ quartz~$\perp$      & &  6.15 & 5.65 & 4.94 & 4.44 & 4.06 & 3.73 & & [\citenum{birch1940thermal}] \\
$\alpha$ quartz~$\parallel$  & & 10.17 & 9.38 & 7.95 & 7.03 & 6.32 & 5.69 & & [\citenum{birch1940thermal}] \\
calcite~$\perp$              & &  3.16 & 3.00 & 2.72 & 2.52 & 2.37 & 2.25 & & [\citenum{birch1940thermal}] \\
calcite~$\parallel$          & &  3.63 & 3.40 & 2.99 & 2.73 & 2.55 & 2.41 & & [\citenum{birch1940thermal}] \\
dolomite                      & &  $4.57^*$ & 4.31 & 3.89 & 3.58 & 3.33 & $3.1^*$   & & [\citenum{birch1940thermal}] \\
albite                     & &  1.58 & 1.51 & 1.46 & 1.42 & 1.41 & 1.4 & & [\citenum{xiong2021thermal}] \\ 
\hline
\end{tabular}
\label{tab:mineral_tc}
\end{table}

\begin{figure}[h!]
  \centering
  \begin{subfigure}[b]{0.45\textwidth}
    \includegraphics[width=\textwidth]{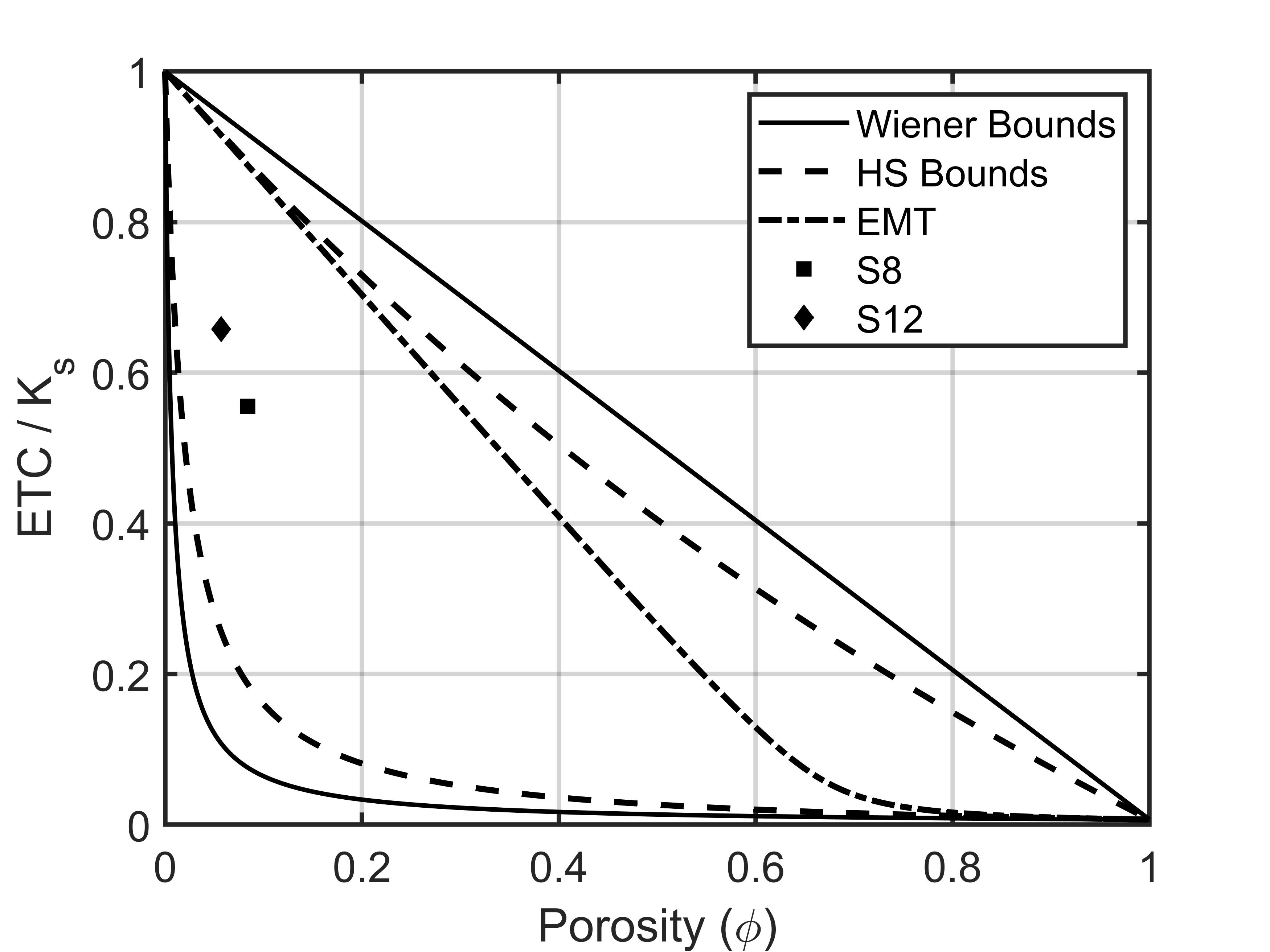}
    \subcaption{}
    \label{fig:BasicETCModels}
  \end{subfigure}
  \hfill
  \begin{subfigure}[b]{0.45\textwidth}
    \includegraphics[width=\textwidth]{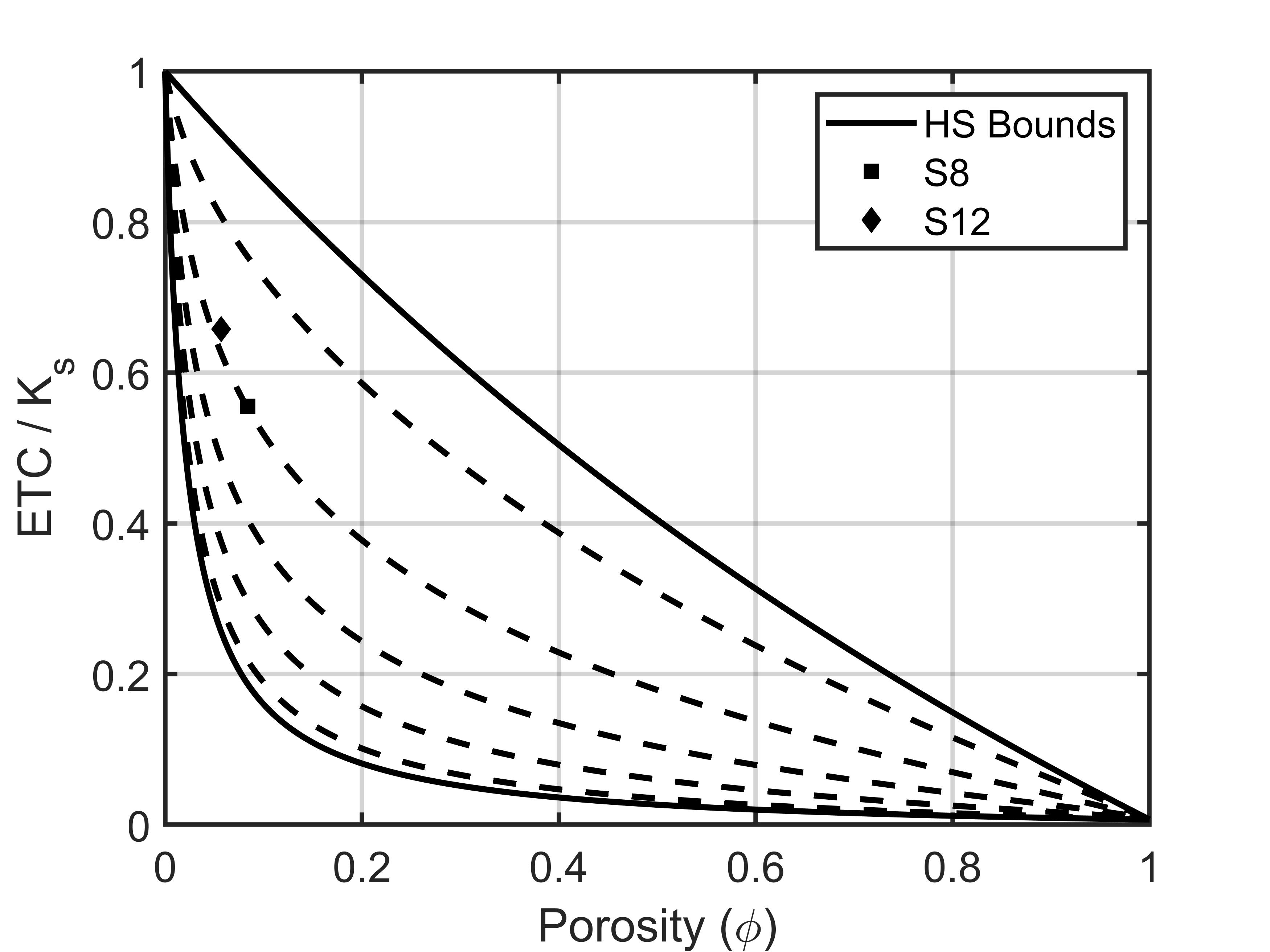}
    \subcaption{}
    \label{fig:CmbinedETCModel}
  \end{subfigure}
  \caption{Comparison of the measured ETC of sandstone at room temperature with model predictions: (a) the five basic structural models (series, parallel, HS upper and lower bounds, and EMT); (b) the weighted geometric mean of HS bounds using various weighting factors: 0.1, 0.3, 0.5, 0.7, and 0.9.}
\end{figure}

Based on the calculated volume fractions and the geometric mean model, the average matrix thermal conductivity for samples S\#8 and S\#12 was estimated to be approximately $\text{k}_{\text{S}} = 3.8~\text{W}~\text{m}^{-1}~\text{K}^{-1}$. The measured values, along with the ETC evaluated from the basic structural models, are presented in Fig.~\ref{fig:BasicETCModels}. Additionally, a comparison between the measured ETC and the combined model based on the geometric weighted mean (Eq.\ref{eq:combi}, with $\text{N} = 0$), constructed using the HN bounds, is presented in Fig.~\ref{fig:CmbinedETCModel}. Although the measured ETC values fall within the calculated bounds, the basic structural models fail to accurately predict the experimental values. This discrepancy is largely attributed to the high contrast between $\text{k}_{\text{f}}$ and $\text{k}_{\text{s}}$ in these models, which is a well-known limitation reported in the literature. However, it can be observed that the weighted combination of the HS bounds offers a significantly improved approximation.

Similar to the experiments, a comparative steady state scheme was used to evaluate the thermal conductivity of the samples in the LEM. A reference layer with the same properties used in the experiments was placed above the sample. Dirichlet boundary conditions were imposed as follows: $\text{T}_1$ at the top of the reference and $\text{T}_3$ at the bottom of the sample. The lateral boundary of the sample was held at $\text{T}_{\text{S}}$, taken as the average of top and bottom boundary conditions. The measurement point $\text{T}_2$ was read at the mid-height of the reference layer. Under steady, one-dimensional conduction with negligible radiation and lateral losses, the sample's thermal conductivity, $\text{k}$, can be evaluated as follows

\begin{equation}
    k = \frac{k_R S_p}{D}; \quad D= S_R\left(\frac{T_1-T_3}{2(T_1-T_2)}-1 \right)
    \label{eq:TC}
\end{equation}

\noindent where $\text{k}_{\text{R}} = 2~\text{W}~\text{m}^{-1}~\text{K}^{-1}$, and $\text{S}_{\text{R}}= 20~\text{mm}$ are thermal conductivity, and height of the reference material, respectively. $\text{S}_\text{p}$ denotes the sample height. Eq.~\ref{eq:TC} holds for ideal one-dimensional conduction with isolated lateral boundaries. In the experiments, however, the side boundary was held at $\text{T}_{\text{s}}$ to reduce losses, which introduces a small but systematic departure from a linear axial profile in the reference–sample stack. Because the LEM reproduces these boundary conditions, a bias correction for the mid-plane reading $\text{T}_2$ was established. Parametric LEM runs for various thermal conductivities of the studied sample were performed, and the deviation of $\text{T}_2$ from its ideal 1D value was evaluated. The normalized bias collapses onto a single curve when expressed as a function of
\begin{equation*}
    R = \frac{T_1-T_3}{T_1-T_2}.
\end{equation*}

\begin{figure}
    \centering
    \includegraphics[width=0.5\linewidth]{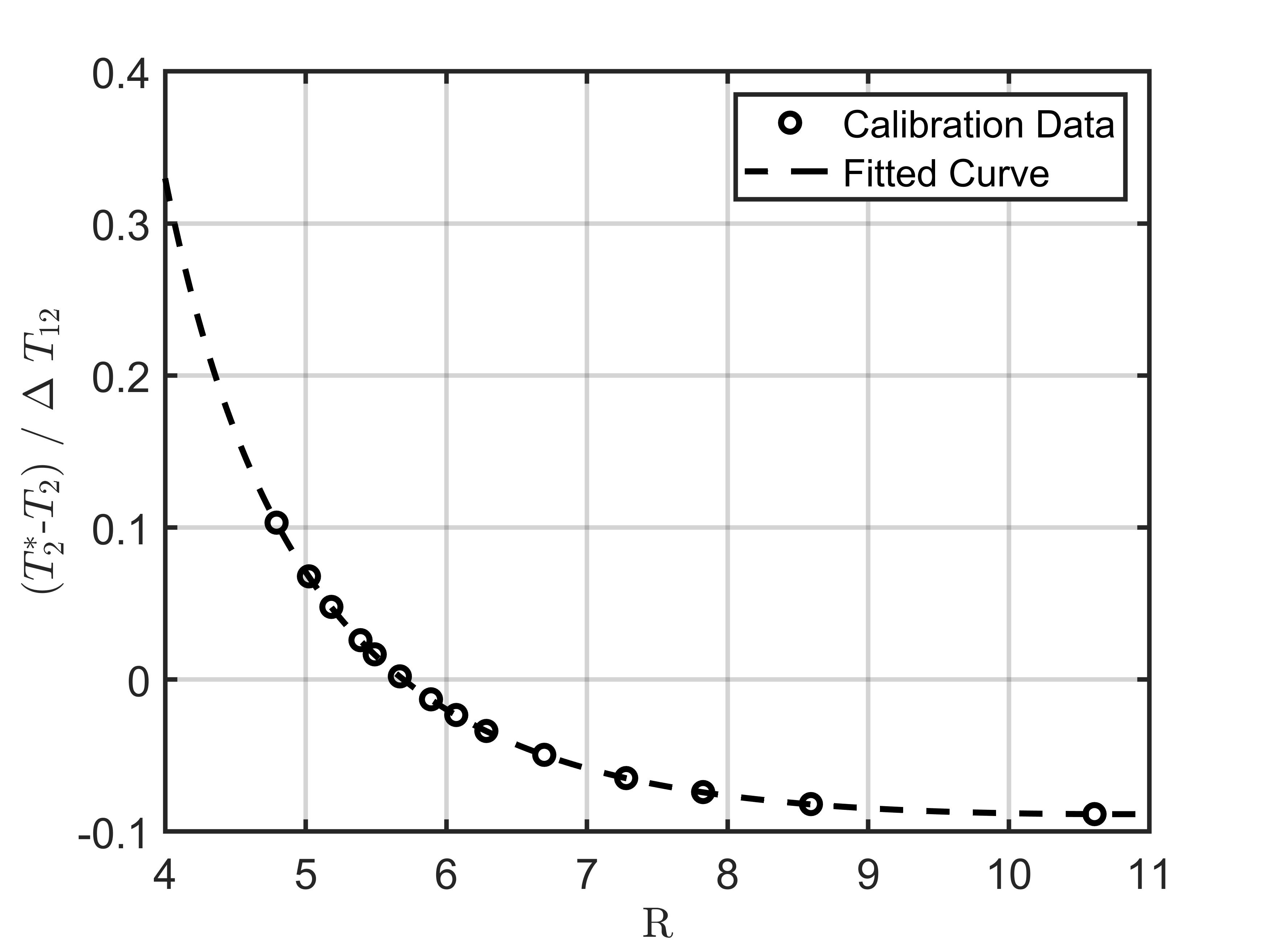}
    \caption{Calibration of the mid-plane sensor bias under guarded side boundaries. Points show LEM calibration data; the dashed line is the fit c(R).}
    \label{fig:LEM_Calibration}
\end{figure}

An empirical fit $\text{c}(\text{R})$ was obtained and used to correct the sensor reading as
\begin{equation}
    T_2^* = T_2 + c(R)\times\Delta T_{12}; \quad \Delta T_{12} = T_1 - T_2
\end{equation}
after which $\text{T}_2^*$ is inserted in Eq.~\ref{eq:TC} for the conductivity evaluation. The calibration data and the fitted function are shown in Fig.~\ref{fig:LEM_Calibration}. The observed error in $\text{T}_2$ stems from the temperature increase in the middle cross-section, which is a consequence of the applied boundary condition $\text{T}_{\text{S}}$. The $\text{T}_{\text{S}}$-induced shift in $\text{T}_2$ readings is solely geometry-dependent.

Before presenting the results, the model ingredients relevant for ambient conditions are summarized. The LEM uses the constituent thermal conductivities (Table \ref{tab:mineral_tc}), elastic properties, fracture toughness, and thermal expansion coefficients together with porosity as model inputs. The full parameter set adopted for the simulations is listed in Table \ref{tab:mech_props}.

\begin{table}[h]
\centering
\caption{Constitutive mechanical parameters used in LEM simulations. Values of minerals' Young's modulus, E (\cite{wang2015elasticity,chen2001elasticity,sayers2008elastic,pabst2015effective}), thermal expansion, $\alpha$ (\cite{siegesmund2011stone}), and fracture toughness, $\text{K}_\text{{IC}}$ (\cite{broz2006microhardness,guo2025fracture}), were obtained from literature.\label{tab:mech_props}}
\begin{tabular}{|l|cccc|}
\hline
Mineral & E (GPa) & $\alpha$ ~$\perp$ ($10^{-6}\,\mathrm{K}^{-1}$) & $\alpha$ ~$\parallel$ ($10^{-6}\,\mathrm{K}^{-1}$) & $\text{K}_{\text{IC}} (\text{MPa}.\text{m}^{1/2})$ \\
\hline
Quartz   & 95.7 & 13.3  & 7.7 &1.6 \\
Calcite  & 86 & -5.6 & 25.1 & 0.39 \\
Dolomite & 123 & 6.2 & 25.8 & 1.86\\
Albite   & 88.8  & 5.6 & 10.5 & 0.88\\
\hline
\end{tabular}
\end{table}

\begin{figure}
    \centering
    \includegraphics[width=0.85\linewidth]{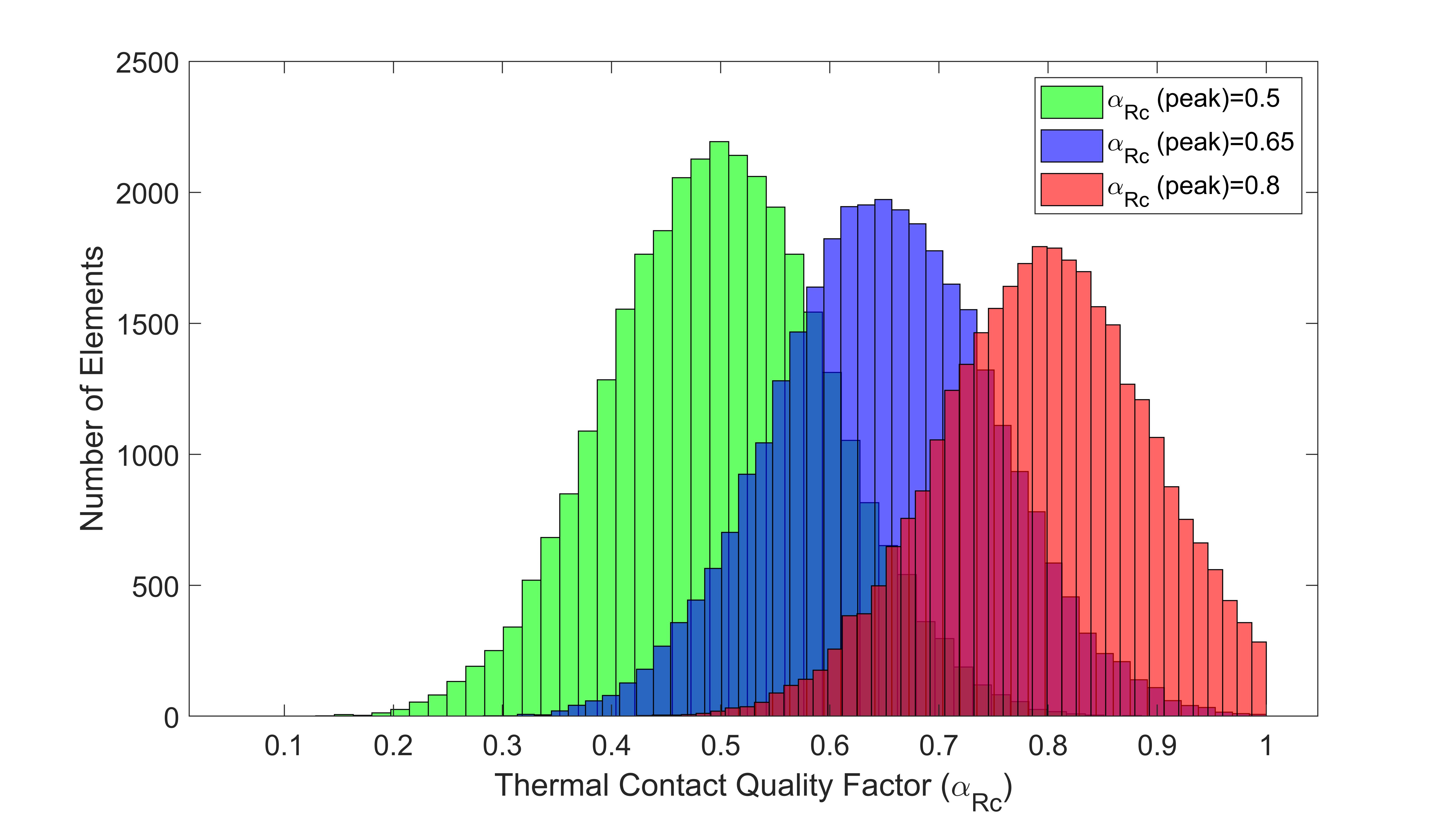}
    \caption{Gaussian distributions of the contact quality factor adopted in the LEM. Three cases represent poor, intermediate, and good grain contacts.}
    \label{fig:contact_quality}
\end{figure}

As mentioned earlier, contacts between grains are represented by a thermal contact quality factor, $\alpha_{\text{Rc}}$. Rather than prescribing a single value, a spatial distribution is assigned to reflect natural variability in grain contacts. A Gaussian distribution was adopted, and three peak values were examined to span poor, intermediate, and good contact scenarios. The resulting histograms are shown in Fig.~\ref{fig:contact_quality}. 

\begin{figure}[h!]
  \centering
  \begin{subfigure}[b]{0.45\textwidth}
    \includegraphics[width=\textwidth]{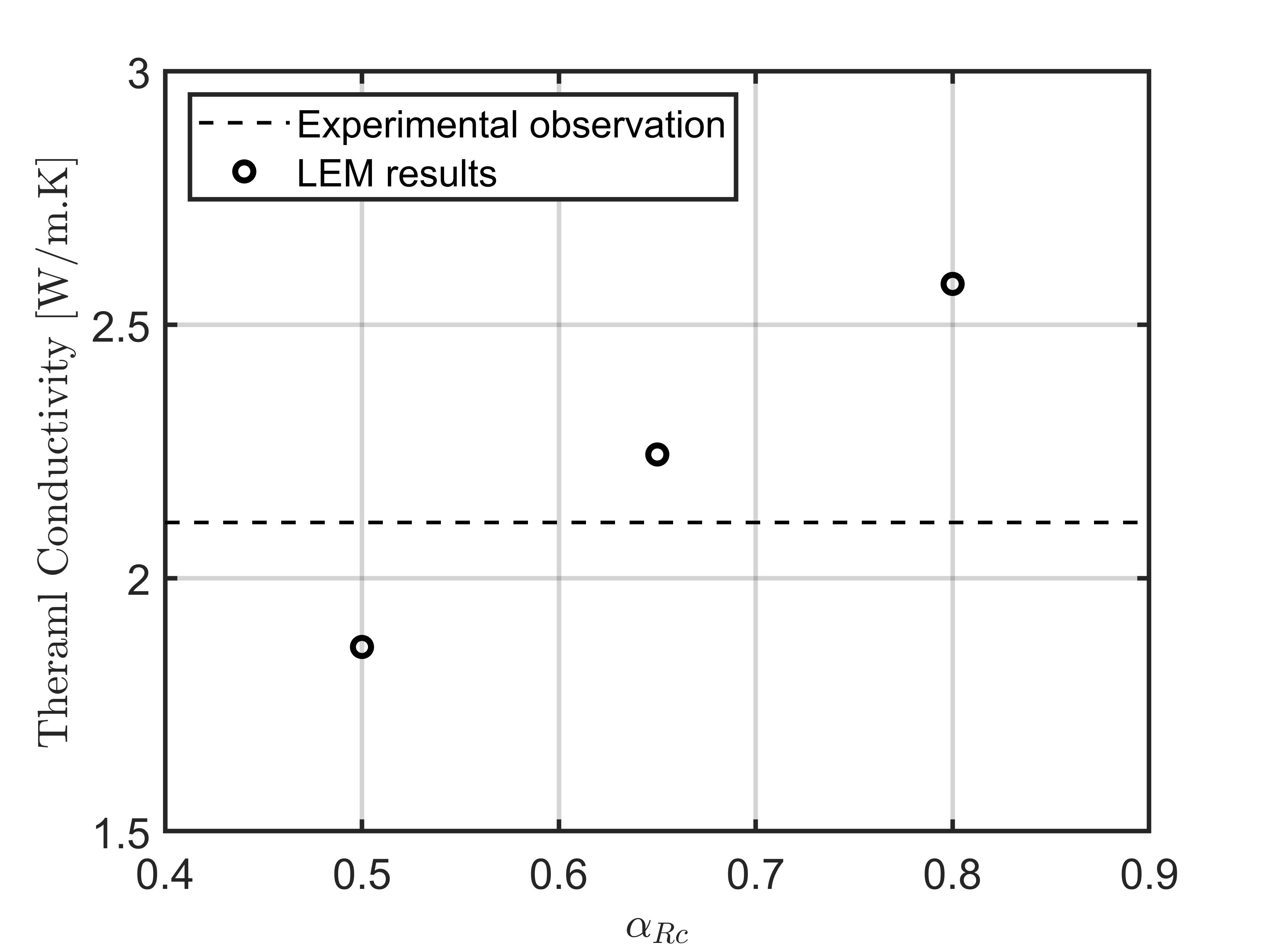}
    \subcaption{}
  \end{subfigure}
  \hfill
  \begin{subfigure}[b]{0.45\textwidth}
    \includegraphics[width=\textwidth]{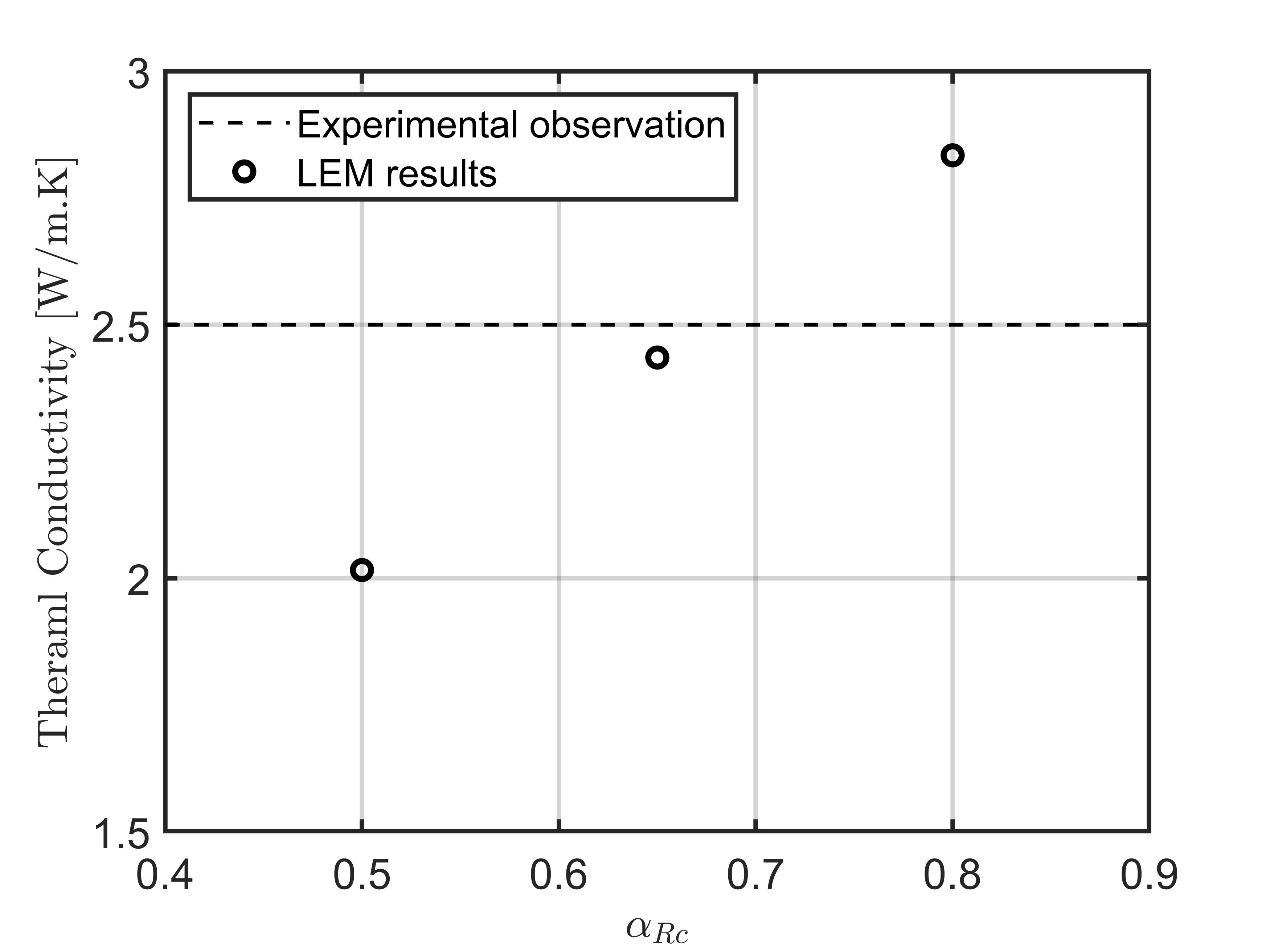}
    \subcaption{}
  \end{subfigure}
  \caption{Comparison of LEM predictions and measured thermal conductivity at room temperature and atmospheric pressure for (a) S\#8 and (b) S\#12.  \label{fig:RoomTemp_LEM_Res}}
\end{figure}

With these assumptions, ambient thermal conductivity at room temperature was computed for S\#8 and S\#12 under three contact quality distributions (Fig. \ref{fig:RoomTemp_LEM_Res}). The aim was to quantify how contact variability influences bulk conductivity at fixed phase fractions and porosity, and to identify the distribution that best reproduces the measurements for each sample. As expected, the predicted conductivity increases with improving contact quality. The best agreement with the measured values is obtained for a contact quality peak of 0.65 in both samples. This distribution is retained for the subsequent pressure–temperature simulations, while the mechanical property set and the critical stress intensity factor remain unchanged.

\begin{figure}
    \centering
    \includegraphics[width=0.5\linewidth]{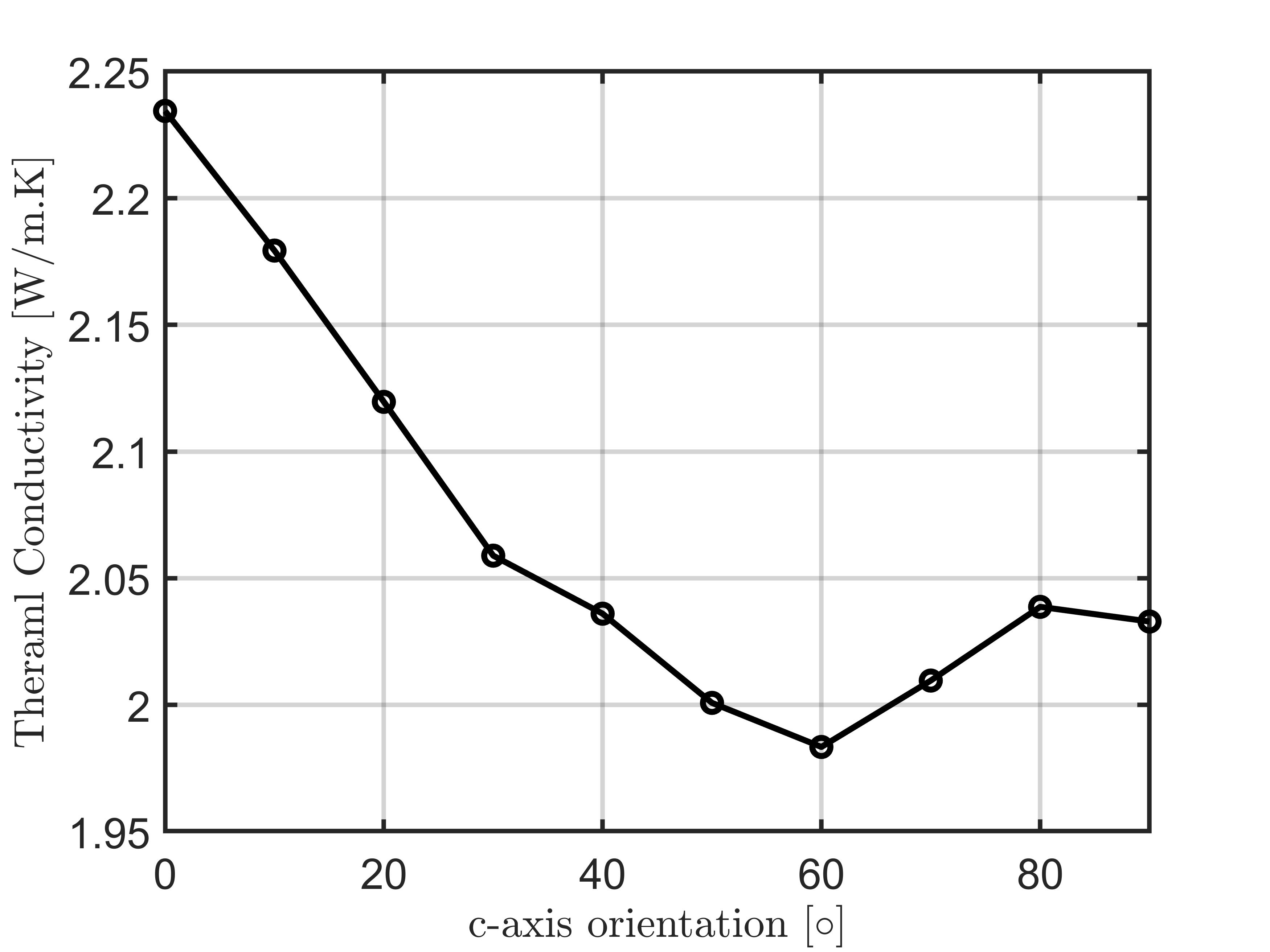}
    \caption{Sensitivity of effective thermal conductivity to crystallographic orientation at room temperature and atmospheric pressure for S\#8. The common c-axis for anisotropic minerals was rotated in ten equal increments from $0~^{\circ}$(vertical-axis) to $90~^{\circ}$ (horizontal-axis).}
    \label{fig:LEM_c_axis}
\end{figure}

It should be noted that the anisotropic thermal conductivities of quartz and calcite were considered in the simulations. For the reported ambient comparisons, a common crystallographic frame was assumed in which the c-axis is aligned with the vertical axis for all grains. To assess the influence of this choice, a deterministic rotation study was performed in which the common c-axis was rotated in ten equal steps from vertical to horizontal. The resulting changes in effective thermal conductivity were small. The aligned c-axis convention is therefore retained for the subsequent analyses. As an example, Fig.~\ref{fig:LEM_c_axis} illustrates the variations of effective thermal conductivity to crystallographic orientation at room temperature and atmospheric pressure for S\#8.

\subsection{Verification under in situ conditions}
\begin{figure}[h!]
  \centering
  \begin{subfigure}[b]{0.45\textwidth}
    \includegraphics[width=\textwidth]{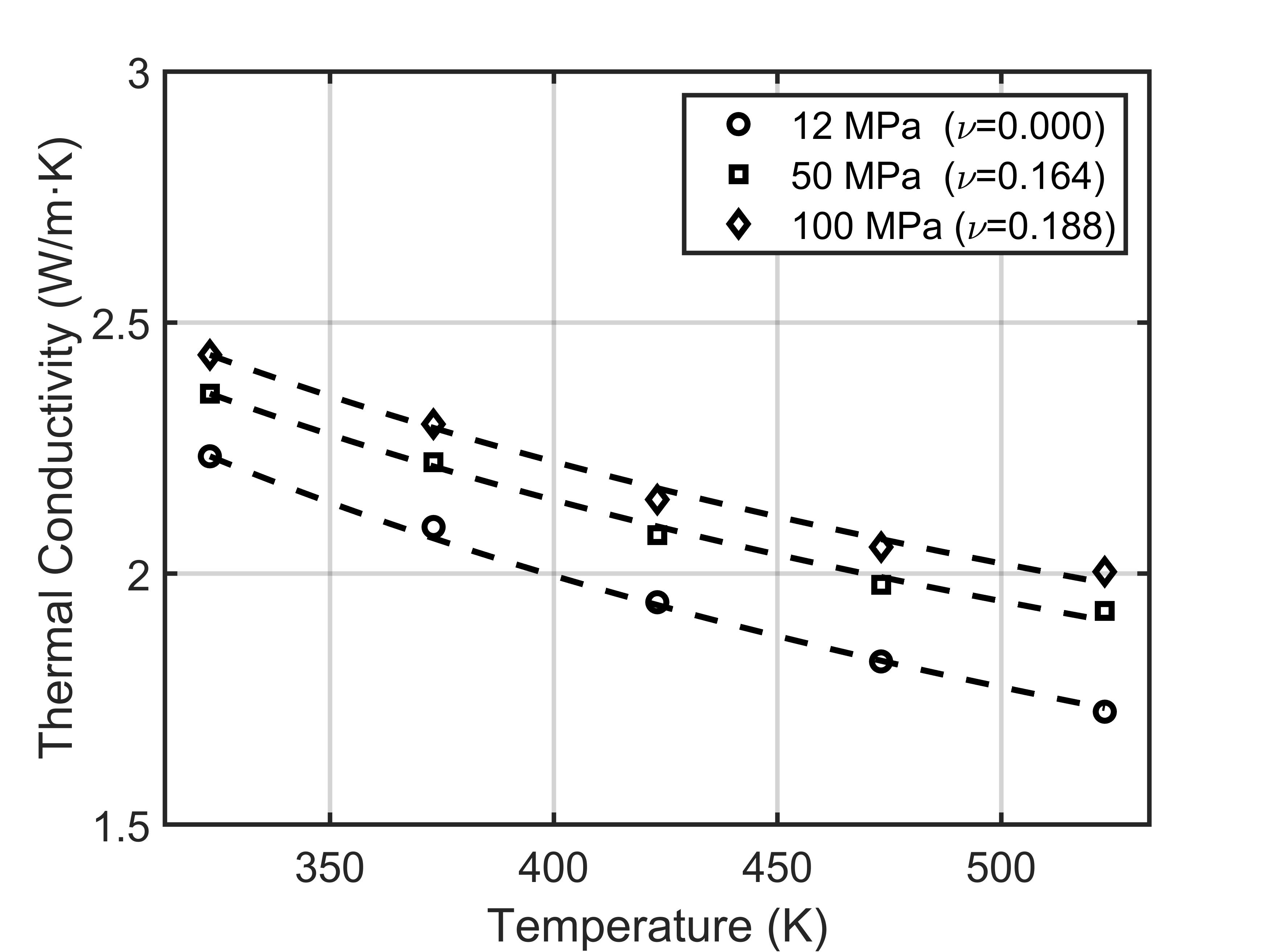}
    \subcaption{}
  \end{subfigure}
  \hfill
  \begin{subfigure}[b]{0.45\textwidth}
    \includegraphics[width=\textwidth]{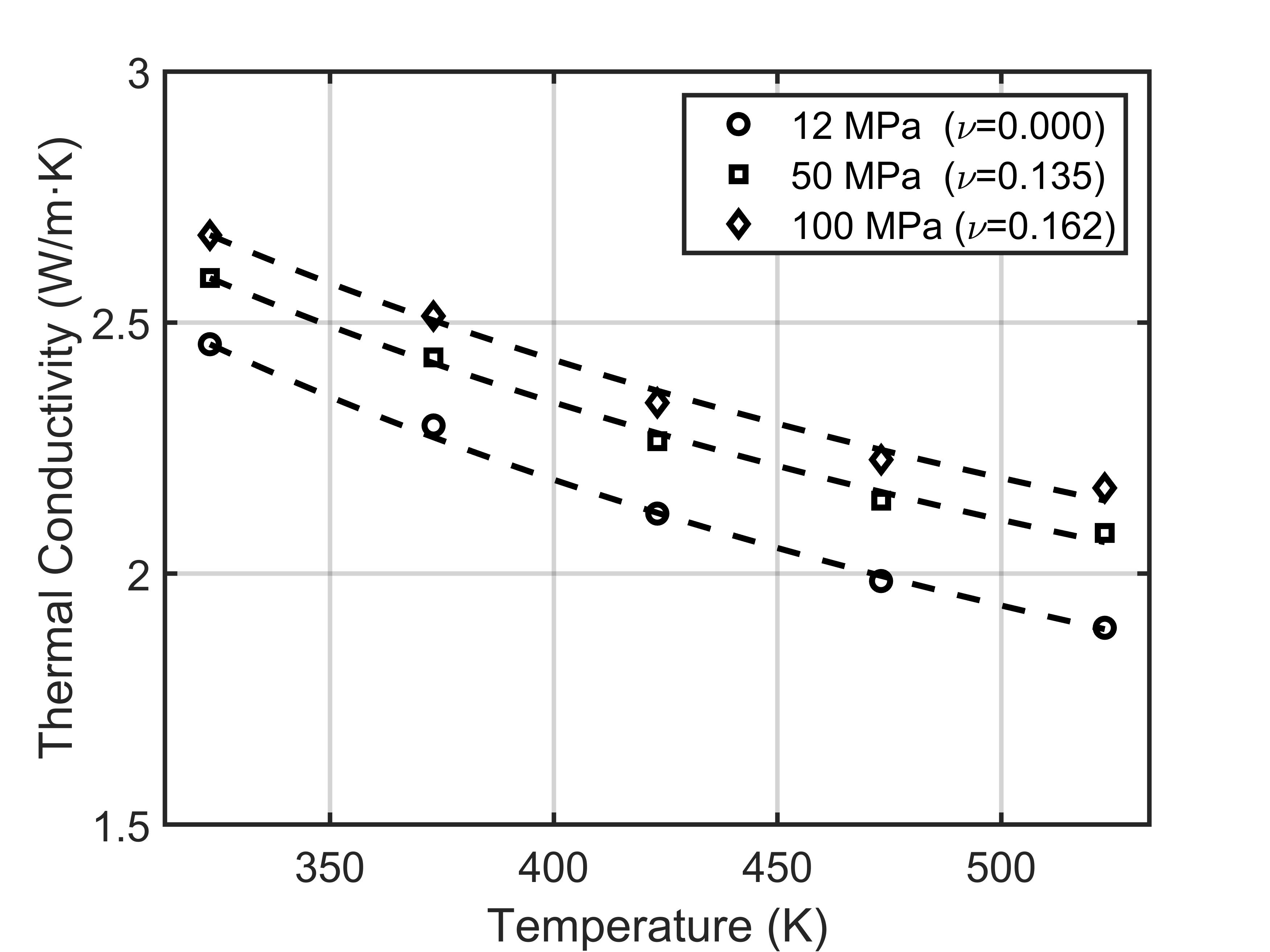}
    \subcaption{}
  \end{subfigure}
  \begin{subfigure}[b]{0.45\textwidth}
    \includegraphics[width=\textwidth]{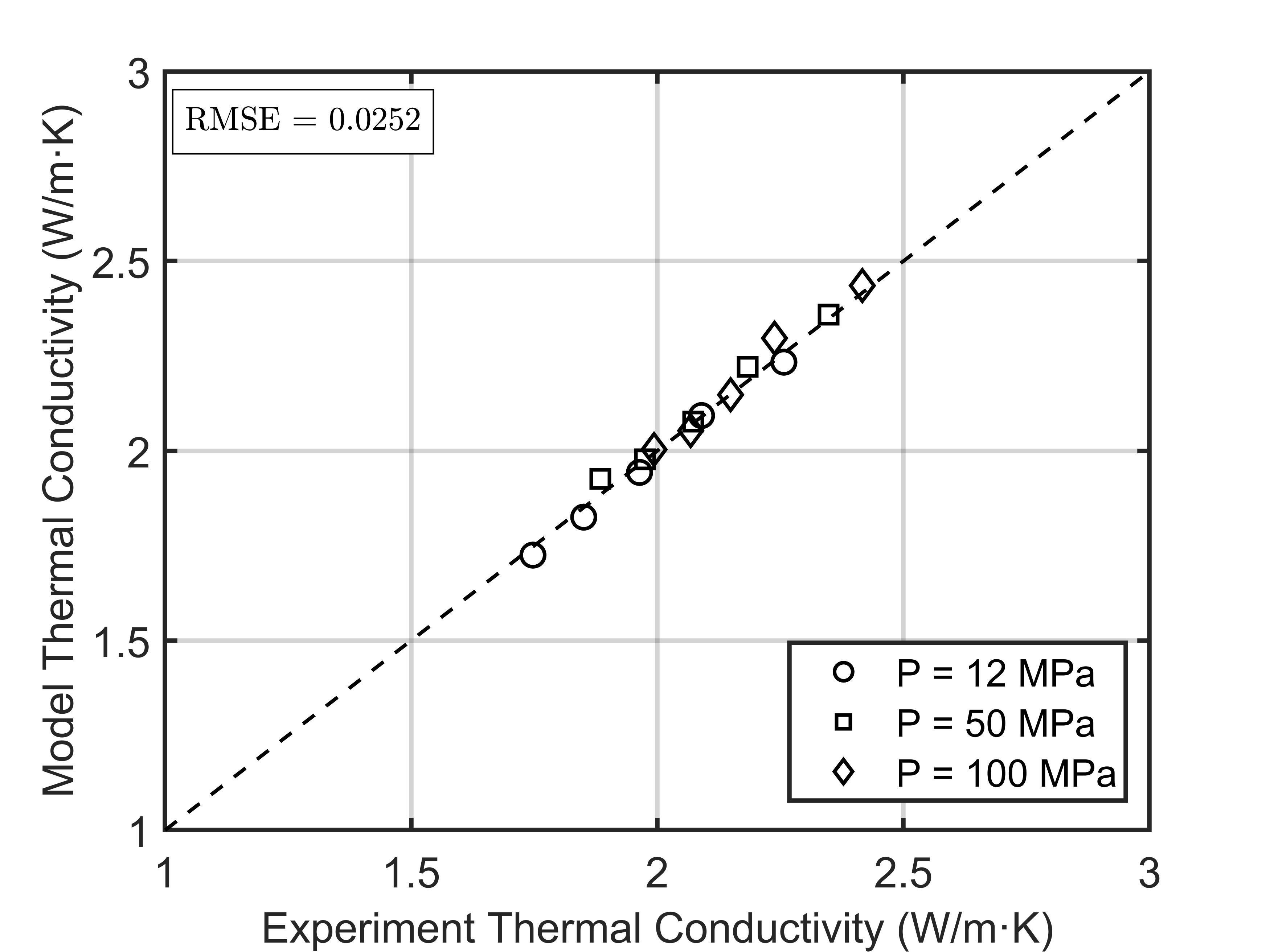}
    \subcaption{}
  \end{subfigure}
  \hfill
  \begin{subfigure}[b]{0.45\textwidth}
    \includegraphics[width=\textwidth]{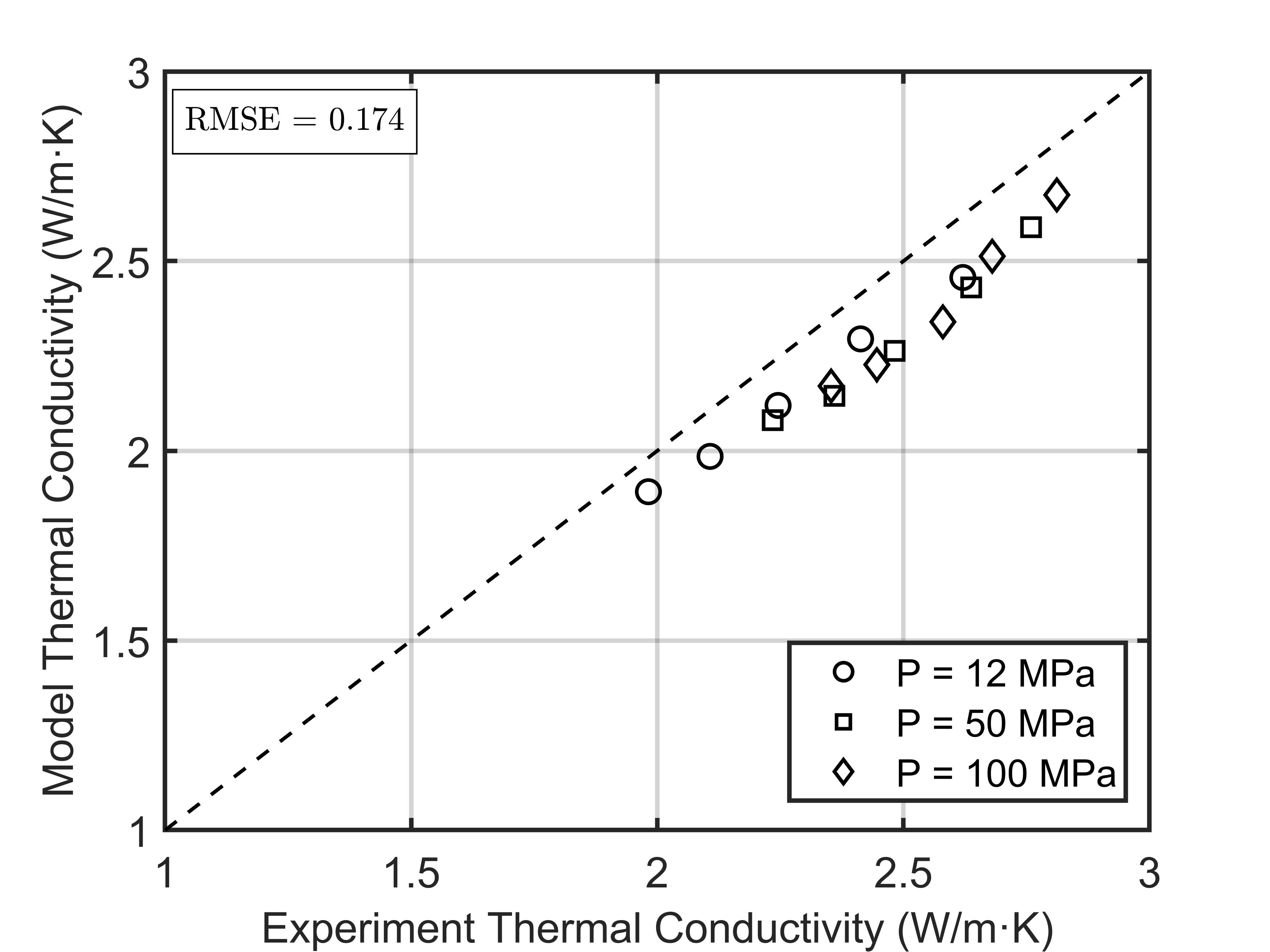}
    \subcaption{}
  \end{subfigure}
  \caption{Comparison of LEM predictions and measurements under in situ conditions for S\#8 (left column) and S\#12 (right column). (a,b) Thermal conductivity versus temperature at 12, 50, and 100 MPa (50–250~$^{\circ}\text{C}$ in 50~$^{\circ}\text{C}$ steps). Markers denote measurements; dashed curves are LEM predictions with fits to Eq.~\ref{eq:lambda_TP}. (c,d) One to one comparison of $\text{k}_{\mathrm{LEM}}$ versus $\text{k}_{\mathrm{exp}}$ under the same pressure-temperature conditions. \label{fig:LEM_EXP_WP}}
\end{figure}

In situ verification was carried out by comparing LEM predictions with the experimental thermal conductivity of S\#8 and S\#12 over the pressure–temperature grid used in the cubic press. All parameters identified at ambient conditions were retained, including the contact quality peak of 0.65, while the mechanical property set and the critical stress intensity factor were kept unchanged. Similar to the applied boundary conditions in the experimental analysis, simulations were performed at confining pressures of 12, 50, and 100 MPa with temperature stepped from $50~^{\circ}\text{C}$ to $250~^{\circ}\text{C}$ in $50~^{\circ}\text{C}$ increments, and results were evaluated against the corresponding measurements. Fig.~\ref{fig:LEM_EXP_WP} summarizes the results for S\#8 and S\#12 in a single view. For each sample, LEM predictions of thermal conductivity across the full pressure–temperature grid are shown together with curves fitted using Eq.~\ref{eq:lambda_TP}. Additionally, a one-to-one comparison of LEM-evaluated thermal conductivity, $\text{k}_{\mathrm{LEM}}$, versus the corresponding measured values, $\text{k}_{\mathrm{exp}}$, is shown. The root mean square error (RMSE) is reported to quantify absolute agreement.

A consistent agreement was achieved with the adopted contact quality coefficient, with S\#8 showing the closest match. The LEM reproduces the expected experimental behavior across the pressure and temperature range. First, the decrease of thermal conductivity with increasing temperature is captured. In theory, this reflects enhanced phonon scattering and the thermally induced opening of compliant features; in the LEM this trend arises primarily from the temperature dependence of the constituent properties. Second, the nonlinear increase of thermal conductivity at lower pressures (below 100 MPa) is reproduced and is attributed to progressive stiffening and enlargement of grain contacts together with the closure of compliant defects, represented in the model through a Hertzian contact law. Third, the mitigating effect of pressure on the temperature-driven decrease is obtained. With increasing temperature, fractures tend to develop along grain boundaries where thermal expansion mismatch concentrates stresses, whereas higher confining pressure suppresses propagation and promotes closure of existing defects, thereby reducing the drop in conductivity.

To illustrate these mechanisms, qualitative maps from the LEM are shown in Fig.~\ref{fig:sand8_frac_profile} for S\#8 at $250~^{\circ}\text{C}$ under two confinement pressures. In the first row, intergranular fractures are displayed; at $12~\text{MPa}$, a denser network is observed, whereas at $100~\text{MPa}$ the fracture set is sparser and dominated by a few short, arrested features. The second row shows the corresponding temperature fields. At $12~\text{MPa}$ the fracture clusters induce more pronounced perturbations of the isotherms, whereas at $100~\text{MPa}$ the isotherms are markedly smoother, consistent with the reduced damage.

\newcommand{\legendraise}{0.6 cm} 
\begin{figure}[h!]
  \centering
  \hspace{-1.2cm}
  \begin{subfigure}[b]{0.45\textwidth}
  \centering
    \includegraphics[scale=0.5]{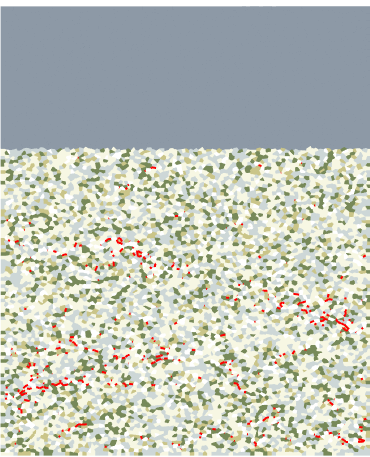}
    \subcaption{}
  \end{subfigure}
  \begin{subfigure}[b]{0.45\textwidth}
  \centering
    \includegraphics[scale=0.5]{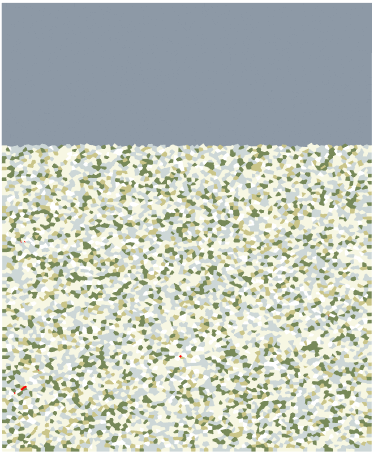}
    \subcaption{}
  \end{subfigure}
  \begin{subfigure}[b]{0.45\textwidth}
  \centering
    \includegraphics[scale=0.5]{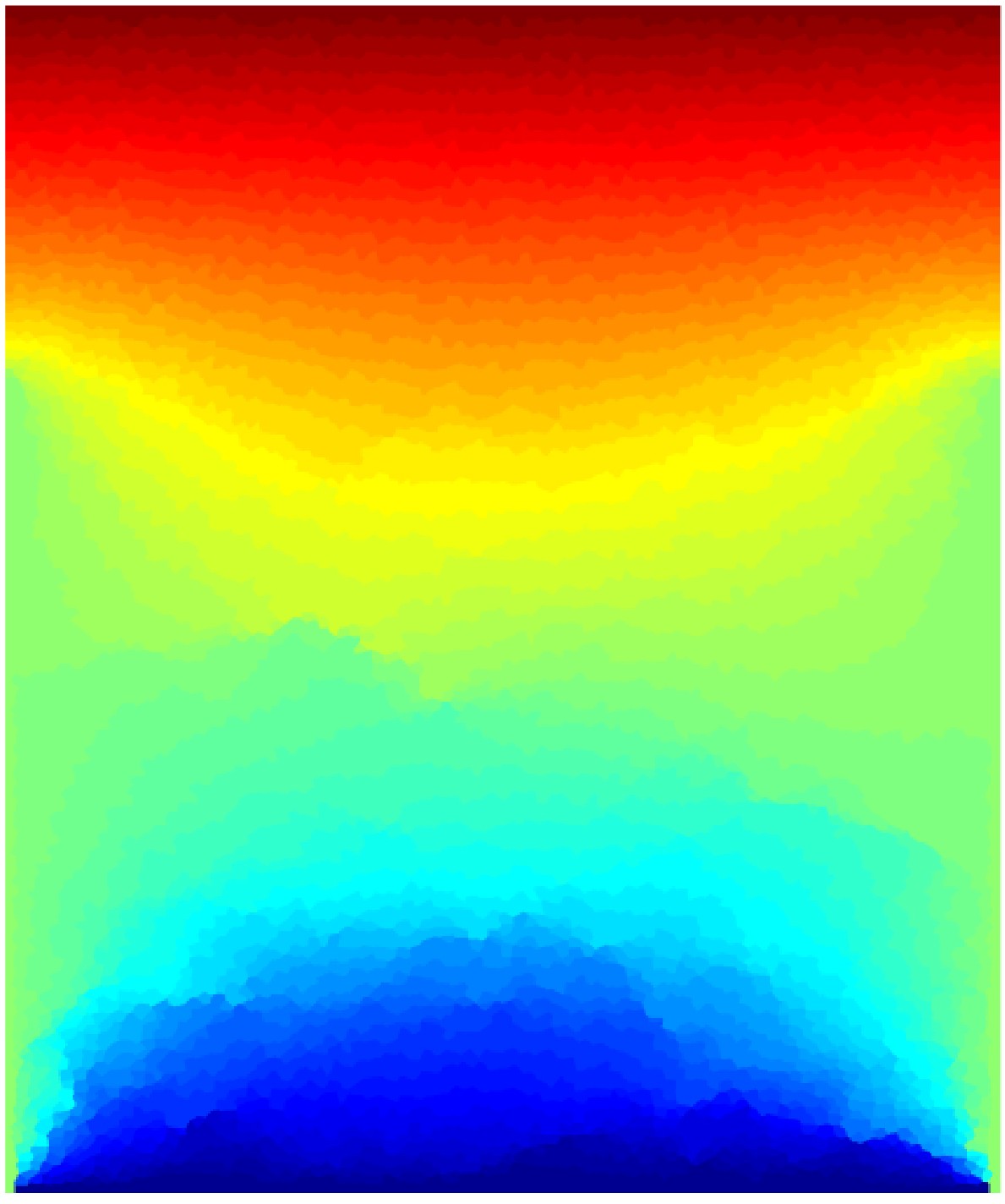}
    \subcaption{}
  \end{subfigure}
  \begin{subfigure}[b]{0.45\textwidth}
  \centering
    \includegraphics[scale=0.5]{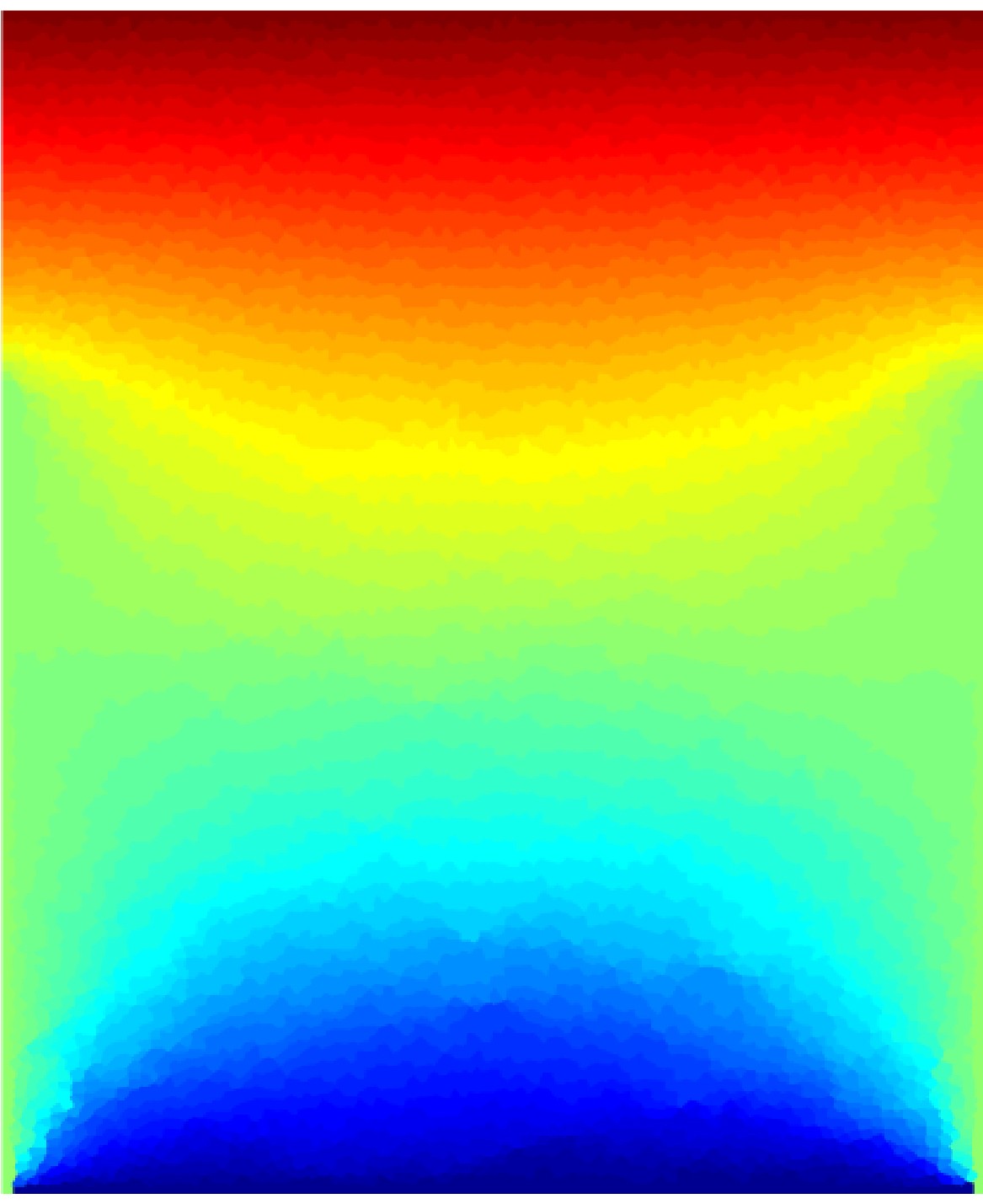}
    \subcaption{}
  \end{subfigure}
  \hspace{-0.8cm}
  \raisebox{\legendraise}[0pt][0pt]{%
    \begin{subfigure}[b]{0.1\textwidth}
      \centering
      \includegraphics[trim=9cm 2cm 0 2cm, clip, scale=0.1]{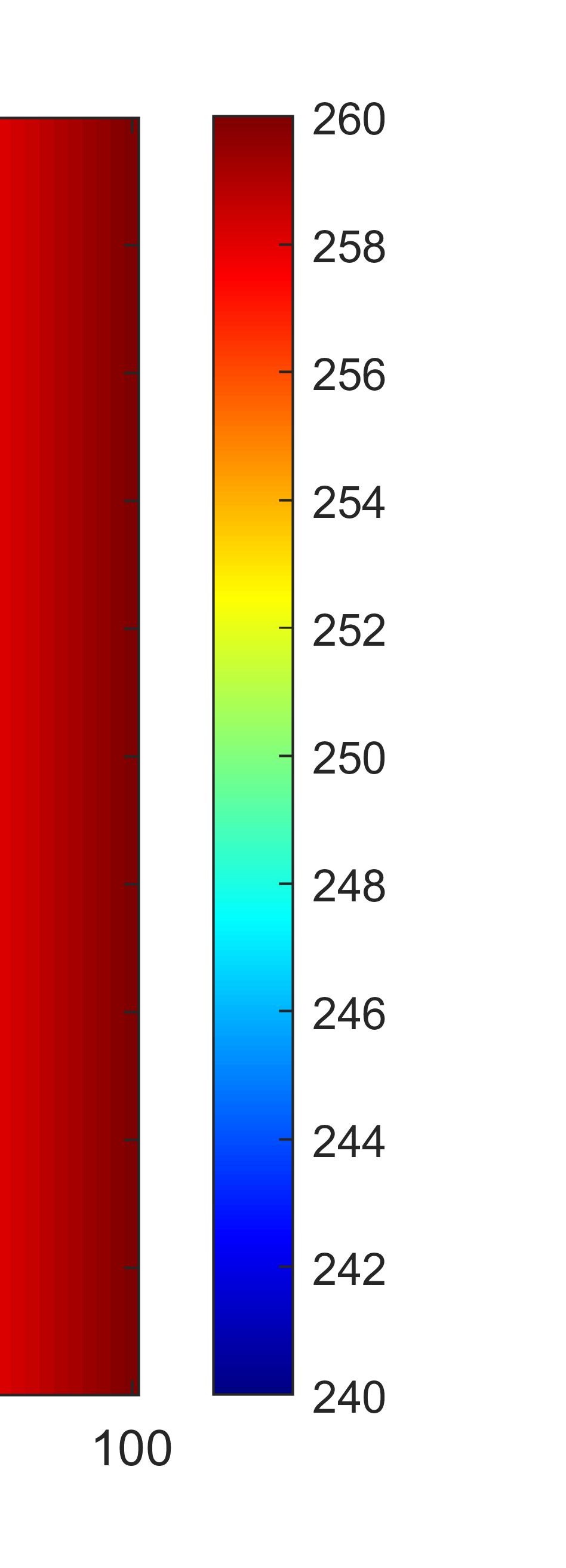}
    \end{subfigure}%
  }
  \caption{Qualitative LEM outputs at $250~^{\circ}\text{C}$ for two confining pressures: (a) fracture map at $12~\text{MPa}$ and (b) at $100~\text{MPa}$, with intergranular fractures shown in red; corresponding temperature fields for (c) $12~\text{MPa}$ and (d) $100~\text{MPa}$. \label{fig:sand8_frac_profile}}
\end{figure}

\section{Conclusion}

The proposed LEM-based methodology provides a microstructure resolving approach to predict rock thermal conductivity under coupled pressure and temperature while explicitly accounting for anisotropic mineral conductivities, intergranular fracturing, and grain contact quality.

Verification was carried out in two steps. First, the LEM response was compared with room temperature and atmospheric pressure measurements of thermal conductivity for two dry sandstones using baseline needle probe tests. From this comparison, the contact quality distribution was identified.  Second, to verify performance under coupled elevated temperature and pressure, the LEM results were compared with measurements on the same samples obtained in a cubic press using a comparative steady state configuration at confining pressures of 12, 50, and 100 MPa and temperatures from 50 to $250~^{\circ}\text{C}$. The expected decrease in thermal conductivity values with temperature, the nonlinear increase at lower confining pressures, and the reduced temperature sensitivity at higher pressures were all captured. Parity analysis and RMSE indicated close agreement for both studied samples. The approach remains practical when only phase fractions and porosity are available, relying on a stochastic volume fraction constrained discretization and a single ambient calibrated contact quality distribution. It clarifies where microstructural resolution adds value beyond mixture rules and provides a validated pathway to predict conductivity across pressure and temperature ranges without detailed imaging.

In the current approach, limitations include reliance on bulk descriptors rather than image-based discretization, and identification of contact quality from ambient data. These aspects could be improved by employing higher-resolution imaging facilities or by selecting samples with greater porosity and mineral contrast, allowing segmentation-based discretization of the actual microstructure. The calibration of grain-contact quality could further benefit from targeted micro-mechanical experiments, e.g., nano-indentation, enabling a more direct link between contact stiffness, surface roughness, and interfacial conductance. Future work should as well focus on developing fully 3D thermo-mechanical models that incorporate matrix anisotropy, supported by numerical parallelization and GPU computing to improve simulation efficiency.
\break
\bibliographystyle{elsarticle-harv} 
\bibliography{References.bib}
\end{document}